\documentclass[aps,prd,groupedaddress,showpacs,superscriptaddress,floatfix]{revtex4-1}

\usepackage{amsmath}
\usepackage{graphicx}
\usepackage{amssymb}
\usepackage{bm}
\usepackage{array}
\usepackage{hyperref}
\usepackage{slashed}
\usepackage{braket}

\begin{document}

\title{Quark orbital motions from Wigner distributions}

\newcommand*{\PKU}{School of Physics and State Key Laboratory of Nuclear Physics and
Technology, Peking University, Beijing 100871,
China}\affiliation{\PKU}

\author{Tianbo Liu}\email{liutb@pku.edu.cn}\affiliation{\PKU}


\begin{abstract}
We investigate quark Wigner distributions in a light-cone spectator model. Both the scalar and the axial-vector spectators are included. The light-cone wave functions are derived from effective quark-spectator-nucleon vertex and then generalized by adjusting the power of energy denominators. The gauge link is taken into account by introducing relative phases to the light-cone amplitudes, and the phases are estimated from one gluon exchange interactions. The mixing distributions, which describe the correlation between transverse coordinate and transverse momentum and represent quark orbital motions, are calculated from the Wigner distributions. We find both $u$ quark and $d$ quark have positive orbital angular momentum in a polarized proton at small $x$ region, but a sign change is observed at large $x$ region for the $d$ quark. Besides, some model relations between Wigner distributions with different polarization configurations are found.
\end{abstract}


\maketitle

\section{Introduction}

Hadrons are bound states of strong interactions which are described by quantum chromodynamics (QCD) in the framework of Yang--Mills gauge theory. Due to the nonperturbative nature of QCD at low energy scale, it is almost impossible to calculate all the properties of hadrons directly from QCD at present. One of the main goals of particle physics is to unravel the quark and gluon structure of hadrons, and it is necessary to investigate the structure of nucleons in details since most experiments on particle physics are based on proton and nucleus beams or targets. The parton model formulated by Feynman and formalized by Bjorken and Paschos~\cite{Feynman:1969ej,Bjorken:1969ja} was proved successful in explaining the high energy hadronic scattering experiments. Based on the parton model and factorizations~\cite{Collins:1987pm}, parton distribution functions (PDFs) are defined as general process independent functions to describe the light-cone longitudinal momentum fraction carried by each flavor of partons. The PDFs play an important role in analyzing the data from high energy hadronic scatterings. With experimental techniques of polarizing targets and beams, one is able to get a richer picture of partonic structures, and spin-related PDFs, such as helicities and transversities, are defined. Including the information of parton transverse distributions, transverse momentum dependent parton distributions (TMDs)~\cite{Collins:1981uk,Collins:1981uw,Kotzinian:1994dv,Mulders:1995dh,Sivers:1989cc,Boer:1997nt} and generalized parton distributions (GPDs)~\cite{Mueller:1998fv,Goeke:2001tz,Diehl:2003ny,Ji:2004gf,Belitsky:2005qn,Boffi:2007yc} are introduced to provide three-dimensional images of hadrons. The TMDs contain transverse momentum distributions of partons, and the GPDs contain transverse coordinate distributions of partons through the impact parameter dependent densities (IPDs). Both of them are useful tools in analying experimental data and understanding the nucleon structures.

As a further generalization, the Wigner distributions are defined as coordinate--momentum joint distributions to understand the partonic structure of a nucleon. The Wigner distribution is a quantum phase-space distribution first introduced by Wigner~\cite{Wigner:1932eb}. It has been applied to many areas of physics, such as the quantum information, quantum molecular dynamics, nonlinear dynamics and optics~\cite{Balazs:1983hk,Hillery:1983ms,Lee1995147}, and is even directly measurable in some experiments~\cite{Vogel:1989zz,Smithey:1993zz,Breitenbach:1997nature,Banaszek:1999ya}. In QCD, the Wigner distribution was first explored as a six-dimensional function~\cite{Ji:2003ak,Belitsky:2003nz} where the nonrelativistic approximation was applied. Then a five-dimensional Wigner distribution is proposed~\cite{Lorce:2011kd} in the light-cone framework or the infinite momentum frame where the parton model is defined. By integrations over transverse coordinates or transverse momentum, the Wigner distributions will reduce to TMDs and IPDs respectively. However, constrained by the Heisenberg uncertainty principle in quantum theories~\cite{aHeisenberg:1927zz}, one cannot know two non-commutable quantities simultaneously. Therefore, we have no probability interpretations for Wigner distributions where the transverse coordinates and transverse momenta do not commute, though one may try to find some certain situations where semiclassical interpretations are possible. Apart from TMDs and IPDs, one can also define mixing distributions by intergrating Wigner distributions over one transverse coordinate and one transverse momentum along two orthogonal directions. The mixing distributions do have the probability interpretations since the remaining transverse coordinate and momentum along two orthogonal directions are commutable. They represent the correlations between transverse coordinate and momentum, and thus the orbital motions of partons can be clearly seen from them.

By combining the polarization configurations (unpolarized, longitudinal polarized and transverse polarized) of the quark and the nucleon, one can define ten independent quark Wigner distributions. In this paper, we investigate all these ten distributions in a light-cone spectator model with both the scalar and the axial-vector spectators included. The Wilson line, {\it i.e.} the gauge link, which plays an important role in time-reversal odd TMDs, are taken into account by introducing relative phases to the light-cone amplitudes. This paper is organized as follows. In Sect. II, we derive the light-cone wave functions in the spectator model, and then calculate the Wigner distributions in Sect. III. We provide the numerical results for unpolarized and longitudinal polarized mixing distributions as well as the orbital angular momentum and the spin-orbit correlator in Sect. IV. Some conclusions are drawn in the last section.

\section{Light-cone wave function in the spectator model}

In the front form of relativistic dynamics~\cite{Dirac:1949cp}, fields are quantized at fixed light-cone time $\tau=(t+z)/\sqrt{2}$ instead of the ordinary time $t$ in the instant form. Hadrons, as bound states of QCD, are eigenstates of the light-cone hamiltonian $H_\textrm{LC}=2P^+P^--\bm{P}_\perp^2$, and the eigenvalues are invariant mass square. One of the advantage of light-cone quantization in QCD is the simple vacuum. Since the light-cone longitudinal momentum $k^+=(k^0+k^3)/\sqrt{2}$ of massive particles is positive definite, the Fock state vacuum $|0\rangle$, {\it i.e.} the free vacuum, is exactly the physical vacuum, if the possibility of the color-singlet states built on massless gluons with zero momentum is ignored~\cite{Brodsky:1997de}. Thus, one may have unambiguous definition of the constituents of hadrons. Then a hadron state can be expanded on a complete Fock state basis as
\begin{equation}
\begin{split}
|\Psi:P^+,\bm{P}_\perp,S_z\rangle=\sum_{n,\{\lambda_i\}}\prod\int\frac{dx_i}{2\sqrt{x_i}}\frac{d^2\bm{k}_{\perp i}}{(2\pi)^3}16\pi^3\delta(1-\sum_ix_i)\delta^{(2)}(\bm{P}_\perp-\sum_i\bm{k}_{\perp i})\\
\times\psi^{S_z}_{n\{\lambda_i\}}(\{x_i\},\{\bm{k}_{\perp i}\})|n:\{x_i\},\{\bm{k}_{\perp i}\},\{\lambda_i\}\rangle,
\end{split}
\end{equation}
where $n$ represents the components of the state, $x_i$ and $\bm{k}_{\perp i}$ are longitudinal momentum fraction $k_i^+/P^+$ and transverse momentum of the $i$th constituent, and $S_z$ and $\lambda_i$ are light-cone helicities of the hadron and the constituent respectively. The $\psi_n$ is the light-cone wave function which describes the probability amplitude of the Fock state $|n\rangle$ in a hadron state, and in principle it should be derived from the QCD lagrangian. However, the QCD is not well understood and it is still challenging to derive the wave functions directly from the lagrangian, though in recent years some attempts were proposed to obtain the first approximation of the light-cone wave functions, for instance the AdS/CFT correspondence between the string state in anti de Sitter (AdS) space and conformal field theories (CFTs) in physical spacetime~\cite{Maldacena:1997re} in front form~\cite{Brodsky:2003px,Brodsky:2006uqa,deTeramond:2008ht}.

In the spectator model, the nucleon is viewed as a struck quark and a spectator which contains the remaining constituents. This model is already applied to investigate the structure functions~\cite{Feynman:1973xc}, form factors~\cite{Ma:2002ir,Ma:2002xu,Liu2014}, TMDs~\cite{Jakob:1997wg,Brodsky:2002cx,She:2009jq,Bacchetta:2008af} and GPDs~\cite{Burkardt:2003je,Chakrabarti:2005zm}. The amplitudes of the quark-spectator states in a nucleon are described by the light-cone wave functions, but as stated in the first section, we cannot directly derive them from QCD lagrangian at present. Many phenomenological light-cone wave functions for spin average quark-spectator state were explored, such as the Brodsky--Huang--Lepage (BHL) prescription~\cite{Brodsky:1980vj,Brodsky:1981jv,Brodsky:1982nx}, the Teren’ev-Karmanov (TK) prescription~\cite{Terentev:1976jk,Karmanov:1979if}, the Chung-Coester-Polyzou (CCP) prescription~\cite{Chung:1988mu} and the Vega--Schmidt--Gutsche--Lyubovitskij (VSGL) prescription~\cite{Vega:2013bxa}. Here we introduce an effective quark-spectator-nucleon coupling to derive the probability amplitude perturbatively~\cite{Jakob:1997wg,Bacchetta:2008af}, and then generalize the form of the wave function by adjusting the power of the energy denominator, {\it i.e.} the propagator~\cite{Hwang:2007tb}.

\begin{figure}
\includegraphics[width=0.3\textwidth]{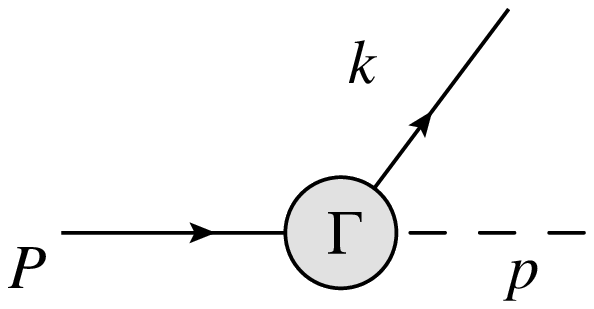}
\includegraphics[width=0.42\textwidth]{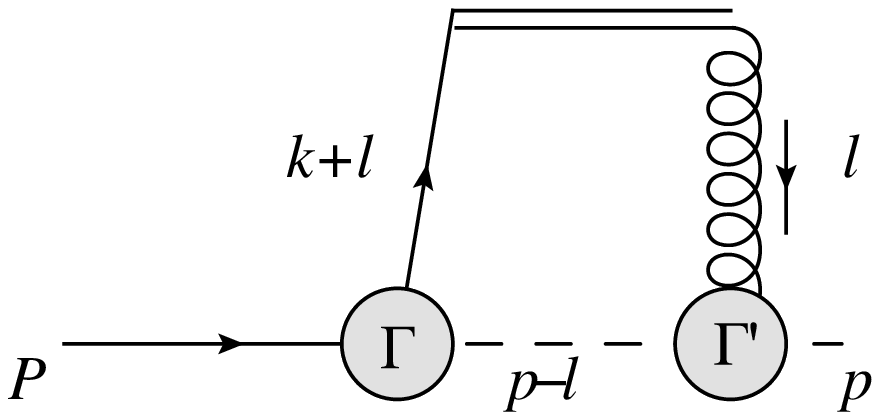}
\caption{Feynman diagrams to calculate the light-cone wave functions. They are drawn with JaxoDraw~\cite{Binosi:2008ig}.\label{feyn}}
\end{figure}

Constained by the quantum numbers of the quark and nucleon, the spectator can be either a scalar or an axial-vector, and the axial-vector one is necessary for flavor separation. Therefore, we introduce two effective interaction terms in the lagrangian as
\begin{equation}
\mathcal{L}_\textrm{I}=-g_s\phi\bar{\psi}\Psi-\frac{g_v}{\sqrt{3}}A_\mu\bar{\psi}\gamma^\mu\gamma_5\Psi+h.c.,
\end{equation}
where $\psi$, $\Psi$, $\phi$ and $A_\mu$ are quark, nuleon, scalar and axial-vector spectator fields. The $g_s$ and $g_v$ are coupling constants, and one may also introduce some suitable form factors to them~\cite{Jakob:1997wg,Bacchetta:2008af}. Then, the amplitude, {\it i.e.} the light-cone wave function， can be calculated either from Feynman rules as in Fig. \ref{feyn} or from light-cone time ordered perturbative theory:
\begin{eqnarray}
\psi^{\Lambda(s)}_\lambda&=&\frac{g_s}{(2\pi)^\frac{3}{2}}\sqrt{\frac{x}{1-x}}\frac{\bar{u}(k,\lambda)U(P,\Lambda)}{[(P-p)^2-m^2]^n},\\
\psi^{\Lambda(v)}_{\lambda\lambda'}&=&\frac{g_v}{\sqrt{3}(2\pi)^\frac{3}{2}}\sqrt{\frac{x}{1-x}}\frac{\bar{u}(k,\lambda)\slashed{\epsilon}^*(p,\lambda')\gamma_5U(P,\Lambda)}{[(P-p)^2-m^2]^n},
\end{eqnarray}
where $u(k,\lambda)$ and $U(P,\Lambda)$ are Dirac spinors for the quark and the nucleon respectively. For $n=1$ case, the amplitudes are derived from perturbative calculations. Then $n$ is introduced to generalize them by adjusting the power behavior~\cite{Hwang:2007tb}. This procedure respects the Lorentz symmetry and it can be induced from a form factor $[(p-p)^2-m^2]^{-(n-1)}$. According to theoretically inspired phenomenological model counting rules and perturbative analysis for the asymptotic drop off of form factors~\cite{Drell:1969km,West:1970av,Brodsky:1974vy}, it is suggested to set $n=2$. Therefore, we only display the light-cone wave functions with $n=2$ here, and one can easily write down the expressions with any other $n$ values. 

We adopt the Brodsky--Lepage convension~\cite{Lepage:1980fj} for light-cone spinors as
\begin{equation}
u(k,\uparrow)=\frac{1}{\sqrt{2^{3/2}k^+}}\left(\begin{array}{c}
\sqrt{2}k^++m\\
k^1+ik^2\\
\sqrt{2}k^+-m\\
k^1+ik^2
\end{array}\right),
\quad
u(k,\downarrow)=\frac{1}{\sqrt{2^{3/2}k^+}}\left(\begin{array}{c}
-k^1+ik^2\\
\sqrt{2}k^++m\\
k^1-ik^2\\
-\sqrt{2}k^++m
\end{array}\right).
\end{equation}
To prepare for the calculations in the next section, we choose the frame as
\begin{eqnarray}
P&=&\bigg(P^+,P^-,-\frac{\bm{\Delta}_\perp}{2}\bigg),\\
k&=&\bigg(xP^+,k^-,\bm{k}_\perp-\frac{\bm{\Delta}_\perp}{2}\bigg),\\
p&=&\bigg((1-x)P^+,p^-,-\bm{k}_\perp\bigg).
\end{eqnarray}
Then, the light-cone wave functions with scalar spectator are
\begin{eqnarray}
\psi^{\uparrow(s)}_\uparrow&=&\frac{g_s}{(2\pi)^\frac{3}{2}}\frac{(1-x)^{\frac{3}{2}}(m+xM)}{[(\bm{k}_\perp-\frac{1-x}{2}\bm{\Delta}_\perp)^2+L_s^2]^2},\\
\psi^{\uparrow(s)}_\downarrow&=&-\frac{g_s}{(2\pi)^\frac{3}{2}}\frac{(1-x)^{\frac{3}{2}}[k^1+ik^2-\frac{1-x}{2}(\Delta^1+i\Delta^2)]}{[(\bm{k}_\perp-\frac{1-x}{2}\bm{\Delta}_\perp)^2+L_s^2]^2},\\
\psi^{\downarrow(s)}_\uparrow&=&\frac{g_s}{(2\pi)^\frac{3}{2}}\frac{(1-x)^{\frac{3}{2}}[k^1-ik^2-\frac{1-x}{2}(\Delta^1-i\Delta^2)]}{[(\bm{k}_\perp-\frac{1-x}{2}\bm{\Delta}_\perp)^2+L_s^2]^2},\\
\psi^{\downarrow(s)}_\downarrow&=&\frac{g_s}{(2\pi)^\frac{3}{2}}\frac{(1-x)^{\frac{3}{2}}(m+xM)}{[(\bm{k}_\perp-\frac{1-x}{2}\bm{\Delta}_\perp)^2+L_s^2]^2},
\end{eqnarray}
where
\begin{equation}
L_s^2=xM_s^2+(1-x)m^2-x(1-x)M^2,
\end{equation}
and $m$, $M_s$ and $M$ are the masses of the quark, spectator and nucleon respectively. For the axial-vector spectator, we adopt the transverse polarization vector as
\begin{eqnarray}
\epsilon(p,+)&=&\bigg(0,-\frac{p^1+ip^2}{\sqrt{2}p^+},-\frac{1}{\sqrt{2}},-\frac{i}{\sqrt{2}}\bigg),\\
\epsilon(p,-)&=&\bigg(0,\frac{p^1-ip^2}{\sqrt{2}p^+},\frac{1}{\sqrt{2}},-\frac{i}{\sqrt{2}}\bigg),
\end{eqnarray}
and for a massive axial-vector spectator we also need to introduce the longitudinal polarization vector
\begin{equation}
\epsilon(p,0)=\bigg(\frac{p^+}{M_v},\frac{\bm{p}_\perp^2-M_v^2}{2M_vp^+},\frac{p^1}{M_v},\frac{p^2}{M_v}\bigg),
\end{equation}
where $M_v$ is the mass of the axial-vector spectator. Then, the light-cone wave functions with axial-vector spectator are
\begin{eqnarray}
\psi^{\uparrow(v)}_{\uparrow+}&=&\frac{g_v}{\sqrt{3}(2\pi)^\frac{3}{2}}\frac{\sqrt{2}(1-x)^{\frac{1}{2}}[k^1-ik^2-\frac{1-x}{2}(\Delta^1-i\Delta^2)]}{[(\bm{k}_\perp-\frac{1-x}{2}\bm{\Delta}_\perp)^2+L_v^2]^2},\\
\psi^{\uparrow(v)}_{\uparrow0}&=&\frac{g_v}{\sqrt{3}(2\pi)^\frac{3}{2}}\frac{(1-x)^\frac{1}{2}[(\bm{k}_\perp-\frac{1-x}{2}\bm{\Delta}_\perp)^2-xM_v^2-(1-x)^2mM]}{M_v[(\bm{k}_\perp-\frac{1-x}{2}\bm{\Delta}_\perp)^2+L_v^2]^2},\\
\psi^{\uparrow(v)}_{\uparrow-}&=&-\frac{g_v}{\sqrt{3}(2\pi)^\frac{3}{2}}\frac{\sqrt{2}(1-x)^\frac{1}{2}x[k^1+ik^2-\frac{1-x}{2}(\Delta^1+i\Delta^2)]}{[(\bm{k}_\perp-\frac{1-x}{2}\bm{\Delta}_\perp)^2+L_v^2]^2},\\
\psi^{\uparrow(v)}_{\downarrow+}&=&\frac{g_v}{\sqrt{3}(2\pi)^\frac{3}{2}}\frac{\sqrt{2}(1-x)^\frac{3}{2}(m+xM)}{[(\bm{k}_\perp-\frac{1-x}{2}\bm{\Delta}_\perp)^2+L_v^2]^2},\\
\psi^{\uparrow(v)}_{\downarrow0}&=&\frac{g_v}{\sqrt{3}(2\pi)^\frac{3}{2}}\frac{(1-x)^\frac{3}{2}(m+M)[k^1+ik^2-\frac{1-x}{2}(\Delta^1+i\Delta^2)]}{M_v[(\bm{k}_\perp-\frac{1-x}{2}\bm{\Delta}_\perp)^2+L_v^2]^2},\\
\psi^{\uparrow(v)}_{\downarrow-}&=&0,\\
\psi^{\downarrow(v)}_{\uparrow+}&=&0,\\
\psi^{\downarrow(v)}_{\uparrow0}&=&\frac{g_v}{\sqrt{3}(2\pi)^\frac{3}{2}}\frac{(1-x)^\frac{3}{2}(m+M)[k^1-ik^2-\frac{1-x}{2}(\Delta^1-i\Delta^2)]}{M_v[(\bm{k}_\perp-\frac{1-x}{2}\bm{\Delta}_\perp)^2+L_v^2]^2},\\
\psi^{\downarrow(v)}_{\uparrow-}&=&-\frac{g_v}{\sqrt{3}(2\pi)^\frac{3}{2}}\frac{\sqrt{2}(1-x)^\frac{3}{2}(m+xM)}{[(\bm{k}_\perp-\frac{1-x}{2}\bm{\Delta}_\perp)^2+L_v^2]^2},\\
\psi^{\downarrow(v)}_{\downarrow+}&=&-\frac{g_v}{\sqrt{3}(2\pi)^\frac{3}{2}}\frac{\sqrt{2}(1-x)^\frac{1}{2}x[k^1-ik^2-\frac{1-x}{2}(\Delta^1-i\Delta^2)]}{[(\bm{k}_\perp-\frac{1-x}{2}\bm{\Delta}_\perp)^2+L_v^2]^2},\\
\psi^{\downarrow(v)}_{\downarrow0}&=&-\frac{g_v}{\sqrt{3}(2\pi)^\frac{3}{2}}\frac{(1-x)^\frac{1}{2}[(\bm{k}_\perp-\frac{1-x}{2}\bm{\Delta}_\perp)^2-xM_v^2-(1-x)^2mM]}{M_v[(\bm{k}_\perp-\frac{1-x}{2}\bm{\Delta}_\perp)^2+L_v^2]^2},\\
\psi^{\downarrow(v)}_{\downarrow-}&=&\frac{g_v}{\sqrt{3}(2\pi)^\frac{3}{2}}\frac{\sqrt{2}(1-x)^{\frac{1}{2}}[k^1+ik^2-\frac{1-x}{2}(\Delta^1+i\Delta^2)]}{[(\bm{k}_\perp-\frac{1-x}{2}\bm{\Delta}_\perp)^2+L_v^2]^2},
\end{eqnarray}
where
\begin{equation}
L_v^2=xM_v^2+(1-x)m^2-x(1-x)M^2.
\end{equation}
As seen from the explicit expressions of the light-cone wave functions, they only depend on the longitudinal momentum fraction $x$ and the intrinsic transverse momentum $\tilde{\bm{k}}_\perp=\bm{k}_\perp-(1-x)\bm{\Delta}_\perp/2$ which are boost invariant variables. Therefore, the light-cone wave functions are frame independent.

\section{Wigner distributions}

In this section, we investigate quark Wigner distributions is the spectator model and the calculations are performed analytically. The numerical results are provided in the next section. 

As a quantum phase space distribution first introduce by Wigner~\cite{Wigner:1932eb}, it contains the most general one-body information in a nucleon. Similar as the quark correlation operator, one can define a Hermitian Wigner operator for quarks at fixed light-cone time as~\cite{Lorce:2011kd}
\begin{equation}\label{wo}
\hat{W}^{[\Gamma]}(x,\bm{b}_\perp,\bm{k}_\perp)=\frac{1}{2}\int\frac{dz^-d^2\bm{z}_\perp}{(2\pi)^3}e^{ik\cdot z}\bar{\psi}(y-\frac{z}{2})\Gamma\mathcal{L}[y-\frac{z}{2},y+\frac{z}{2}]\psi(y+\frac{z}{2})|_{z^+=0},
\end{equation}
where $y=(0,0,\bm{b}_\perp)$ and $\Gamma$ is a twist-two Dirac $\gamma$-matrix $\gamma^+$, $\gamma^+\gamma_5$ or $i\sigma^{j+}\gamma_5$ which corresponds unpolarized, longitudial polarized and transverse polarized quark respectively. The $\mathcal{L}$ is the gauge link connecting quark fields at two points $y-z/2$ and $y+z/2$ to ensure the SU(3) color gauge invariance of the Wigner operator. The Wilson line, {\it i.e.} the gauge link, plays an important role in studying the time-reversal odd TMDs~\cite{Brodsky:2002cx,Collins:2004nx}. The path of the gauge link is chosen as
\begin{equation}
(0,\frac{z^-}{2},\bm{b}_\perp-\frac{\bm{z}_\perp}{2})\rightarrow(0,\infty,\bm{b}_\perp-\frac{\bm{z}_\perp}{2})\rightarrow(0,\infty,\bm{b}_\perp+\frac{\bm{z}_\perp}{2})\rightarrow(0,\frac{z^-}{2},\bm{b}_\perp+\frac{\bm{z}_\perp}{2}),
\end{equation}
in order to obtain the appropriate Wilson line when taking the TMD and IPD limits. By interpolating the Wigner operator (\ref{wo}) into initial and final nucleon state with a momentum $\bm{\Delta}_\perp$ transferred, one can define the Wigner distribution as
\begin{equation}
\rho^{[\Gamma]}(x,\bm{b}_\perp,\bm{k}_\perp,\bm{S})=\int\frac{d^2\bm{\Delta}_\perp}{(2\pi)^2}\left\langle P',\bm{S}\right|\hat{W}^{[\Gamma]}(x,\bm{b}_\perp,\bm{k}_\perp)\left|P,\bm{S}\right\rangle,
\end{equation}
where
\begin{eqnarray}
P&=&\left(P^+,\frac{M^2+\frac{\bm{\Delta}_\perp^2}{4}}{2P^+},-\frac{\bm{\Delta}_\perp}{2}\right),\\
P'&=&\left(P^+,\frac{M^2+\frac{\bm{\Delta}_\perp^2}{4}}{2P^+},\frac{\bm{\Delta}_\perp}{2}\right),
\end{eqnarray}
and $\bm{S}$ is the spin of the nucleon. With the combination of the polarization configurations, unpolarized (U), longitudinal polarized (L) and transverse polartized (T), of the quark and the nucleon, we can define ten independent Wigner distributions: the unpolarized Wigner distribution
\begin{equation}
\rho_{_\textrm{UU}}(x,\bm{b}_\perp,\bm{k}_\perp)=\frac{1}{2}\big[\rho^{[\gamma^+]}(x,\bm{b}_\perp,\bm{k}_\perp,\hat{e}_z)+\rho^{[\gamma^+]}(x,\bm{b}_\perp,\bm{k}_\perp,-\hat{e}_z)\big],
\end{equation}
the unpol-longitudial Wigner distribution
\begin{equation}
\rho_{_\textrm{UL}}(x,\bm{b}_\perp,\bm{k}_\perp)=\frac{1}{2}\big[\rho^{[\gamma^+\gamma_5]}(x,\bm{b}_\perp,\bm{k}_\perp,\hat{e}_z)+\rho^{[\gamma^+\gamma_5]}(x,\bm{b}_\perp,\bm{k}_\perp,-\hat{e}_z)\big],
\end{equation}
the unpol-transverse Wigner distribution
\begin{equation}
\rho_{_\textrm{UT}}^j(x,\bm{b}_\perp,\bm{k}_\perp)=\frac{1}{2}\big[\rho^{[i\sigma^{j+}\gamma_5]}(x,\bm{b}_\perp,\bm{k}_\perp,\hat{e}_z)+\rho^{[i\sigma^{j+}\gamma_5]}(x,\bm{b}_\perp,\bm{k}_\perp,-\hat{e}_z)\big],
\end{equation}
the longi-unpolarized Wigner distribution
\begin{equation}
\rho_{_\textrm{LU}}(x,\bm{b}_\perp,\bm{k}_\perp)=\frac{1}{2}\big[\rho^{[\gamma^+]}(x,\bm{b}_\perp,\bm{k}_\perp,\hat{e}_z)-\rho^{[\gamma^+]}(x,\bm{b}_\perp,\bm{k}_\perp,-\hat{e}_z)\big],
\end{equation}
the longitudinal Wigner distribution
\begin{equation}
\rho_{_\textrm{LL}}(x,\bm{b}_\perp,\bm{k}_\perp)=\frac{1}{2}\big[\rho^{[\gamma^+\gamma_5]}(x,\bm{b}_\perp,\bm{k}_\perp,\hat{e}_z)-\rho^{[\gamma^+\gamma_5]}(x,\bm{b}_\perp,\bm{k}_\perp,-\hat{e}_z)\big],
\end{equation}
the longi-transverse Wigner distribution
\begin{equation}
\rho_{_\textrm{LT}}^j(x,\bm{b}_\perp,\bm{k}_\perp)=\frac{1}{2}\big[\rho^{[i\sigma^{j+}\gamma_5]}(x,\bm{b}_\perp,\bm{k}_\perp,\hat{e}_z)-\rho^{[i\sigma^{j+}\gamma_5]}(x,\bm{b}_\perp,\bm{k}_\perp,-\hat{e}_z)\big],
\end{equation}
the trans-unpolarized Wigner distribution
\begin{equation}
\rho_{_\textrm{TU}}^i(x,\bm{b}_\perp,\bm{k}_\perp)=\frac{1}{2}\big[\rho^{[\gamma^+]}(x,\bm{b}_\perp,\bm{k}_\perp,\hat{e}_i)-\rho^{[\gamma^+]}(x,\bm{b}_\perp,\bm{k}_\perp,-\hat{e}_i)\big],
\end{equation}
the trans-longitudinal Wigner distribution
\begin{equation}
\rho_{_\textrm{TL}}^i(x,\bm{b}_\perp,\bm{k}_\perp)=\frac{1}{2}\big[\rho^{[\gamma^+\gamma_5]}(x,\bm{b}_\perp,\bm{k}_\perp,\hat{e}_i)-\rho^{[\gamma^+\gamma_5]}(x,\bm{b}_\perp,\bm{k}_\perp,-\hat{e}_i)\big],
\end{equation}
the transverse Wigner distribution
\begin{equation}
\rho_{_\textrm{TT}}(x,\bm{b}_\perp,\bm{k}_\perp)=\frac{1}{2}\delta_{ij}\big[\rho^{[i\sigma^{j+}\gamma_5]}(x,\bm{b}_\perp,\bm{k}_\perp,\hat{e}_i)-\rho^{[i\sigma^{j+}\gamma_5]}(x,\bm{b}_\perp,\bm{k}_\perp,-\hat{e}_i)\big],
\end{equation}
and the pretzelous Wigner distribution
\begin{equation}
\rho_{_\textrm{TT}}^\perp(x,\bm{b}_\perp,\bm{k}_\perp)=\frac{1}{2}\epsilon_{ij}\big[\rho^{[i\sigma^{j+}\gamma_5]}(x,\bm{b}_\perp,\bm{k}_\perp,\hat{e}_i)-\rho^{[i\sigma^{j+}\gamma_5]}(x,\bm{b}_\perp,\bm{k}_\perp,-\hat{e}_i)\big],
\end{equation}
where $\delta_{ij}$ is the Kronecker symbol and $\epsilon_{ij}$ is the antisymmetric tensor with $\epsilon_{12}=1$. The names are given by considering the polarization of the quark with a prefix describing the polarization of the nucleon unless they are paralell polarized. The last one is named after the pretzelosity TMD to describe the situation that the quark and nucleon are polarized along two orthogonal transverse directions.

Wigner distributions have direction connection to the generalized parton correlation functions (GPCFs)~\cite{Meissner:2009ww,Meissner:2008ay} which parametrize the fully unintegrated off-diagonal quark-quark correlator. After the integration over the light-cone energy, one obtains the generalized transverse momentum depencent parton distributions (GTMDs), and Wigner distributions can be viewed as transverse Fourier transformation of GTMDs. Unlike the GTMDs which in general are complex-valued functions, Wigner distributions are always real-valued functions. Constrained by the Heisenberg uncertainty principle~\cite{aHeisenberg:1927zz}, we have no probability interpretations for Wigner distributions, though one may still try to find certain situations to have semiclassical interpretations. However, apart from TMDs and IPDs which can be obtained by integrations of Wigner distributions over transverse coordinates and transverse momenta respectively, one can define the mixing distributions which also have probability interpretations by integrating over a transverse coordinate and a transverse momentum along two orthogonal directions:
\begin{equation}
\tilde{\rho}(x,b_x,k_y)=\int db_ydk_x\rho(x,\bm{b}_\perp,\bm{k}_\perp).
\end{equation}
Since the mixing distributions represent the correlation between transverse coordinate and transverse momentum, quark orbital motions are clearly seen from them.

To calculate the distributions, we use the light-cone wave functions derived in Sect. II. The gauge links are taken into account by introducing a phase to each light-cone amplitude~\cite{Brodsky:2002cx} as
\begin{equation}
\mathcal{A}^{\Lambda(s/v)}_{\lambda\lambda'}=\psi^{\Lambda(s/v)}_{\lambda\lambda'}e^{i\varphi^{\Lambda(s/v)}_{\lambda\lambda'}},
\end{equation}
and the phases are estimated from one gluon exchange interactions as in Fig. \ref{feyn}. The gluon spectator coupling vertices are chosen as
\begin{eqnarray}
\Gamma'^{(s)}_\mu&=&ie_c(2p+l)_\mu,\\
\Gamma'^{(v)}_{\alpha\beta\mu}&=&ie_c[(p-\kappa l)_\alpha g_{\beta\mu}+(p+(1+\kappa)l)_\beta g_{\alpha_\mu}-(2p+l)_\mu g_{\alpha\beta}],
\end{eqnarray}
where $e_c$ is the color charge, and $\kappa$ is the anomalous chromomagnetic moment~\cite{Robinson:1986ui,Bacchetta:2008af}. With explicit calculations, we find that each of the phase $\varphi$ is infrared divergent, but the differences between them are infrared finite. Then we remove a infrared divergent phase and keep the finite relative phases as
\begin{equation}\label{amps}
\begin{split}
\mathcal{A}^{\uparrow(s)}_{\uparrow}=\psi^{\uparrow(s)}_{\uparrow},\quad \mathcal{A}^{\uparrow(s)}_{\downarrow}=\psi^{\uparrow(s)}_{\downarrow}e^{i\chi^{(s)}},\\
\mathcal{A}^{\downarrow(s)}_{\uparrow}=\psi^{\downarrow(s)}_{\uparrow}e^{i\chi^{(s)}},\quad \mathcal{A}^{\downarrow(s)}_{\downarrow}=\psi^{\downarrow(s)}_{\downarrow},
\end{split}
\end{equation}
\begin{equation}\label{ampv}
\begin{split}
&\mathcal{A}^{\uparrow(v)}_{\uparrow+}=\psi^{\uparrow(v)}_{\uparrow+}e^{i\chi^{(v)}_a},\quad \mathcal{A}^{\uparrow(v)}_{\uparrow0}=\psi^{\uparrow(v)}_{\uparrow0}e^{i\chi^{(v)}_b}, \quad \mathcal{A}^{\uparrow(v)}_{\uparrow-}=\psi^{\uparrow(v)}_{\uparrow-},\\
&\mathcal{A}^{\uparrow(v)}_{\downarrow+}=\psi^{\uparrow(v)}_{\downarrow+},\quad \mathcal{A}^{\uparrow(v)}_{\downarrow0}=\psi^{\uparrow(v)}_{\downarrow0}e^{i\chi^{(v)}_c},\quad \mathcal{A}^{\uparrow(v)}_{\downarrow-}=0,\\
&\mathcal{A}^{\downarrow(v)}_{\uparrow+}=0,\quad \mathcal{A}^{\downarrow(v)}_{\uparrow0}=\psi^{\downarrow(v)}_{\uparrow0}e^{i\chi^{(v)}_c},\quad \mathcal{A}^{\downarrow(v)}_{\uparrow-}=\psi^{\downarrow(v)}_{\uparrow-},\\
&\mathcal{A}^{\downarrow(v)}_{\downarrow+}=\psi^{\downarrow(v)}_{\downarrow+},\quad \mathcal{A}^{\downarrow(v)}_{\downarrow0}=\psi^{\downarrow(v)}_{\downarrow(0)}e^{i\chi^{(v)}_b},\quad \mathcal{A}^{\downarrow(v)}_{\downarrow-}=\psi^{\downarrow(v)}_{\downarrow-}e^{i\chi^{(v)}_a}.
\end{split}
\end{equation}
Here we display the relative phases for $n=2$ case.
\begin{eqnarray}
\chi^{(s)}&=&\arctan\frac{C_F\alpha_S}{2}\frac{\tilde{\bm{k}}_\perp^2+L_s^2}{L_s^2},\label{phs}\\
\chi^{(v)}_a&=&\arctan\frac{C_F\alpha_S}{2}\frac{(1-x)(\tilde{\bm{k}}_\perp^2+L_v^2)}{L_v^2},\label{phva}\\
\chi^{(v)}_b&=&\arctan\frac{C_F\alpha_S}{2}\frac{(1-x)\tilde{\bm{k}}_\perp^2(\tilde{\bm{k}}_\perp^2+L_v^2)}{L_v^2[\tilde{\bm{k}}_\perp^2-xM_v^2-(1-x)^2mM]},\label{phvb}\\
\chi^{(v)}_c&=&\arctan\frac{C_F\alpha_S}{2}\frac{(1-x)M(\tilde{\bm{k}}_\perp^2+L_v^2)}{L_v^2(m+M)},\label{phvc}
\end{eqnarray}
where $C_F$ is the color factor, $\alpha_S$ is the strong coupling constant and $\tilde{\bm{k}}_\perp$ is the intrinsic transverse momentum which is $\bm{k}_\perp-(1-x)\bm{\Delta}_\perp/2$ for the initial state and $\bm{k}_\perp+(1-x)\bm{\Delta}_\perp/2$ for the final state.

With Eqs. (\ref{amps})-(\ref{phvc}), we can calculate all the distributions. For the scalar spectator,
\begin{align}
&\rho^{(s)}_{_\textrm{UU}}(x,\bm{b}_\perp,\bm{k}_\perp)\nonumber\\
=&\int\frac{d^2\bm{\Delta}_\perp}{(2\pi)^2}\frac{8g_s^2(1-x)^3[4(m+xM)^2\cos(\bm{b}_\perp\cdot\bm{\Delta}_\perp)+(4\bm{k}_\perp^2-(1-x)^2\bm{\Delta}_\perp^2)\cos(\bm{b}_\perp\cdot\bm{\Delta}_\perp+2\delta\chi^{(s)})]}{\pi^3[(4\bm{k}_\perp^2+4L_s^2+(1-x)^2\bm{\Delta}_\perp^2)^2-16(1-x)^2(\bm{k}_\perp\cdot\bm{\Delta}_\perp)^2]^2},\label{uu}\\
&\rho^{(s)}_{_\textrm{UL}}(x,\bm{b}_\perp,\bm{k}_\perp)
=\int\frac{d^2\bm{\Delta}_\perp}{(2\pi)^2}\frac{32g_s^2(1-x)^4\bm{k}_\perp\times\bm{\Delta}_\perp\sin(\bm{b}_\perp\cdot\bm{\Delta}_\perp+2\delta\chi^{(s)})}{\pi^3[(4\bm{k}_\perp^2+4L_s^2+(1-x)^2\bm{\Delta}_\perp^2)^2-16(1-x)^2(\bm{k}_\perp\cdot\bm{\Delta}_\perp)^2]^2},\label{uls}\\
&\rho^{(s)}_{_\textrm{LU}}(x,\bm{b}_\perp,\bm{k}_\perp)
=\int\frac{d^2\bm{\Delta}_\perp}{(2\pi)^2}\frac{32g_s^2(1-x)^4\bm{\Delta}_\perp\times\bm{k}_\perp\sin(\bm{b}_\perp\cdot\bm{\Delta}_\perp+2\delta\chi^{(s)})}{\pi^3[(4\bm{k}_\perp^2+4L_s^2+(1-x)^2\bm{\Delta}_\perp^2)^2-16(1-x)^2(\bm{k}_\perp\cdot\bm{\Delta}_\perp)^2]^2},\label{lus}\\
&\rho^{(s)}_{_\textrm{LL}}(x,\bm{b}_\perp,\bm{k}_\perp)\nonumber\\
=&\int\frac{d^2\bm{\Delta}_\perp}{(2\pi)^2}\frac{8g_s^2(1-x)^3[4(m+xM)^2\cos(\bm{b}_\perp\cdot\bm{\Delta}_\perp)-(4\bm{k}_\perp^2-(1-x)^2\bm{\Delta}_\perp^2)\cos(\bm{b}_\perp\cdot\bm{\Delta}_\perp+2\delta\chi^{(s)})]}{\pi^3[(4\bm{k}_\perp^2+4L_s^2+(1-x)^2\bm{\Delta}_\perp^2)^2-16(1-x)^2(\bm{k}_\perp\cdot\bm{\Delta}_\perp)^2]^2},\\
&\frac{k^j}{M}\rho^{j(s)}_{_\textrm{UT}}(x,\bm{b}_\perp,\bm{k}_\perp)
=\int\frac{d^2\bm{\Delta}_\perp}{(2\pi)^2}\frac{32g_s^2(1-x)^4\bm{k}_\perp\times\bm{\Delta}_\perp(m+xM)\cos\overline{\chi}^{(s)}\sin(\bm{b}_\perp\cdot\bm{\Delta}_\perp+\delta\chi^{(s)})}{\pi^3M[(4\bm{k}_\perp^2+4L_s^2+(1-x)^2\bm{\Delta}_\perp^2)^2-16(1-x)^2(\bm{k}_\perp\cdot\bm{\Delta}_\perp)^2]^2},\\
&\frac{k^j}{M}\rho^{j(s)}_{_\textrm{LT}}(x,\bm{b}_\perp,\bm{k}_\perp)\nonumber\\
=&\int\frac{d^2\bm{\Delta}_\perp}{(2\pi)^2}\frac{32g_s^2(1-x)^3(m+xM)[(1-x)\bm{k}_\perp\cdot\bm{\Delta}_\perp\sin\overline{\chi}^{(s)}\sin(\bm{b}_\perp\cdot\bm{\Delta}_\perp+\delta\chi^{(s)})-2\bm{k}_\perp^2\cos\overline{\chi}^{(s)}\cos(\bm{b}_\perp\cdot\bm{\Delta}_\perp+\delta\chi^{(s)})]}{\pi^3M[(4\bm{k}_\perp^2+4L_s^2+(1-x)^2\bm{\Delta}_\perp^2)^2-16(1-x)^2(\bm{k}_\perp\cdot\bm{\Delta}_\perp)^2]^2},\\
&\frac{k^j}{M}\rho^{j(s)}_{_\textrm{TU}}(x,\bm{b}_\perp,\bm{k}_\perp)
=\int\frac{d^2\bm{\Delta}_\perp}{(2\pi)^2}\frac{32g_s^2(1-x)^4\bm{k}_\perp\times\bm{\Delta}_\perp(m+xM)\cos\overline{\chi}^{(s)}\sin(\bm{b}_\perp\cdot\bm{\Delta}_\perp+\delta\chi^{(s)})}{\pi^3M[(4\bm{k}_\perp^2+4L_s^2+(1-x)^2\bm{\Delta}_\perp^2)^2-16(1-x)^2(\bm{k}_\perp\cdot\bm{\Delta}_\perp)^2]^2},\\
&\frac{k^j}{M}\rho^{j(s)}_{_\textrm{TL}}(x,\bm{b}_\perp,\bm{k}_\perp)\nonumber\\
=&\int\frac{d^2\bm{\Delta}_\perp}{(2\pi)^2}\frac{32g_s^2(1-x)^3(m+xM)[2\bm{k}_\perp^2\cos\overline{\chi}^{(s)}\cos(\bm{b}_\perp\cdot\bm{\Delta}_\perp+\delta\chi^{(s)})-(1-x)\bm{k}_\perp\cdot\bm{\Delta}_\perp\sin\overline{\chi}^{(s)}\sin(\bm{b}_\perp\cdot\bm{\Delta}_\perp+\delta\chi^{(s)})]}{\pi^3M[(4\bm{k}_\perp^2+4L_s^2+(1-x)^2\bm{\Delta}_\perp^2)^2-16(1-x)^2(\bm{k}_\perp\cdot\bm{\Delta}_\perp)^2]^2},\\
&\rho^{(s)}_{_\textrm{TT}}(x,\bm{b}_\perp,\bm{k}_\perp)
=\int\frac{d^2\bm{\Delta}_\perp}{(2\pi)^2}\frac{64g_s^2(1-x)^3(m+xM)^2\cos(\bm{b}_\perp\cdot\bm{\Delta}_\perp)}{\pi^3[(4\bm{k}_\perp^2+4L_s^2+(1-x)^2\bm{\Delta}_\perp^2)^2-16(1-x)^2(\bm{k}_\perp\cdot\bm{\Delta}_\perp)^2]^2},\\
&\rho^{\perp(s)}_{_\textrm{TT}}(x,\bm{b}_\perp,\bm{k}_\perp)=0,\label{pret}
\end{align}
where
\begin{equation}
\overline{\chi}=\frac{1}{2}(\chi_f+\chi_i),\quad \delta\chi=\frac{1}{2}(\chi_f-\chi_i).
\end{equation}
The subscripts $f$ and $i$ represent final state and initial state respectively. Some relations between these distributions can be found from the expressions Eqs. (\ref{uu})-(\ref{pret}):
\begin{eqnarray}
\rho^{(s)}_{_\textrm{UL}}(x,\bm{b}_\perp,\bm{k}_\perp)&=&-\rho^{(s)}_{_\textrm{LU}}(x,\bm{b}_\perp,\bm{k}_\perp),\label{rel1}\\
\rho^{j(s)}_{_\textrm{UT}}(x,\bm{b}_\perp,\bm{k}_\perp)&=&\rho^{j(s)}_{_\textrm{TU}}(x,\bm{b}_\perp,\bm{k}_\perp),\\
\rho^{j(s)}_{_\textrm{LT}}(x,\bm{b}_\perp,\bm{k}_\perp)&=&-\rho^{j(s)}_{_\textrm{TL}}(x,\bm{b}_\perp,\bm{k}_\perp),\\
\rho^{(s)}_{_\textrm{TT}}(x,\bm{b}_\perp,\bm{k}_\perp)&=&\rho^{(s)}_{_\textrm{UU}}(x,\bm{b}_\perp,\bm{k}_\perp)+\rho^{(s)}_{_\textrm{LL}}(x,\bm{b}_\perp,\bm{k}_\perp).\label{rel4}
\end{eqnarray}

For the axial-vector spectator,
\begin{align}
&\rho^{(v)}_{_\textrm{UU}}(x,\bm{b}_\perp,\bm{k}_\perp)\nonumber\\
=&\int\frac{d^2\bm{\Delta}_\perp}{(2\pi)^2}\frac{16g_v^2}{3\pi^3D^2}(1-x)\bigg\{\big[4(1-x)^2(m+xM)^2+4x^2\bm{k}_\perp^2-x^2(1-x)^2\bm{\Delta}_\perp^2\big]\cos(\bm{b}_\perp\cdot\bm{\Delta}_\perp)\nonumber\\
&\quad\quad +\big[4\bm{k}_\perp^2-(1-x)^2\bm{\Delta}_\perp^2\big]\cos(\bm{b}_\perp\cdot\bm{\Delta}_\perp+2\delta\chi^{(v)}_a)+\frac{1}{8M_v^2}\big[4(1-x)^2(m+M)^2\big(4\bm{k}_\perp^2-(1-x)^2\bm{\Delta}_\perp^2\big)\cos(\bm{b}_\perp\cdot\bm{\Delta}_\perp+2\delta\chi^{(v)}_c)\nonumber\\
&\quad\quad +[(4\bm{k}_\perp^2-4(1-x)^2mM-4xM_v^2+(1-x)^2\bm{\Delta}_\perp)^2-16(1-x)^2(\bm{k}_\perp\cdot\bm{\Delta}_\perp)^2]\cos(\bm{b}_\perp\cdot\bm{\Delta}_\perp+2\delta\chi^{(v)}_b)\big]\bigg\},\label{uuv}\\
&\rho^{(v)}_{_\textrm{UL}}(x,\bm{b}_\perp,\bm{k}_\perp)\nonumber\\
=&\int\frac{d^2\bm{\Delta}_\perp}{(2\pi)^2}\frac{32g_v^2}{3\pi^3D^2}(1-x)^2\bm{k}_\perp\times\bm{\Delta}_\perp\bigg\{2\sin(\bm{b}_\perp\cdot\bm{\Delta}_\perp+2\delta\chi^{(v)}_a)-2x^2\sin(\bm{b}_\perp\cdot\bm{\Delta}_\perp)\nonumber\\
&\quad\quad +\frac{1}{M_v^2}(1-x)^2(m+M)^2\sin(\bm{b}_\perp\cdot\bm{\Delta}_\perp+2\delta\chi^{(v)}_c)\bigg\},\label{ulv}\\
&\rho^{(v)}_{_\textrm{LU}}(x,\bm{b}_\perp,\bm{k}_\perp)\nonumber\\
=&\int\frac{d^2\bm{\Delta}_\perp}{(2\pi)^2}\frac{32g_v^2}{3\pi^3D^2}(1-x)^2\bm{k}_\perp\times\bm{\Delta}_\perp\bigg\{2\sin(\bm{b}_\perp\cdot\bm{\Delta}_\perp+2\delta\chi^{(v)}_a)-2x^2\sin(\bm{b}_\perp\cdot\bm{\Delta}_\perp)\nonumber\\
&\quad\quad -\frac{1}{M_v^2}(1-x)^2(m+M)^2\sin(\bm{b}_\perp\cdot\bm{\Delta}_\perp+2\delta\chi^{(v)}_c)\bigg\},\label{luv}\\
&\rho^{(v)}_{_\textrm{LL}}(x,\bm{b}_\perp,\bm{k}_\perp)\nonumber\\
=&\int\frac{d^2\bm{\Delta}_\perp}{(2\pi)^2}\frac{16g_v^2}{3\pi^3D^2}(1-x)\bigg\{\big[4x^2\bm{k}_\perp^2-x^2(1-x)^2\bm{\Delta}_\perp^2-4(1-x)^2(m+xM)^2\big]\cos(\bm{b}_\perp\cdot\bm{\Delta}_\perp)\nonumber\\
&\quad\quad +\big[4\bm{k}_\perp^2-(1-x)^2\bm{\Delta}_\perp^2\big]\cos(\bm{b}_\perp\cdot\bm{\Delta}_\perp+2\delta\chi^{(v)}_a)-\frac{1}{8M_v^2}\big[4(1-x)^2(m+M)^2\big(4\bm{k}_\perp^2-(1-x)^2\bm{\Delta}_\perp^2\big)\cos(\bm{b}_\perp\cdot\bm{\Delta}_\perp+2\delta\chi^{(v)}_c)\nonumber\\
&\quad\quad -[(4\bm{k}_\perp^2-4(1-x)^2mM-4xM_v^2+(1-x)^2\bm{\Delta}_\perp)^2-16(1-x)^2(\bm{k}_\perp\cdot\bm{\Delta}_\perp)^2]\cos(\bm{b}_\perp\cdot\bm{\Delta}_\perp+2\delta\chi^{(v)}_b)\big]\bigg\},\\
&\frac{k^j}{M}\rho^{j(v)}_{_\textrm{UT}}(x,\bm{b}_\perp,\bm{k}_\perp)\nonumber\\
=&\int\frac{d^2\bm{\Delta}_\perp}{(2\pi)^2}\frac{64g_v^2}{3\pi^3MD^2}(1-x)^3\bm{k}_\perp\times\bm{\Delta}_\perp\bigg\{(m+xM)\cos\overline{\chi}^{(v)}_a\sin(\bm{b}_\perp\cdot\bm{\Delta}_\perp+\delta\chi^{(v)}_a)\nonumber\\
&\quad\quad -\frac{1}{8M_v^2}(m+M)[4(1-x)\bm{k}_\perp\cdot\bm{\Delta}_\perp\sin(\overline{\chi}^{(v)}_b-\overline{\chi}^{(v)}_c)\cos(\bm{b}_\perp\cdot\bm{\Delta}_\perp+\delta\chi^{(v)}_b+\delta\chi^{(v)}_c)\nonumber\\
&\quad\quad +(4\bm{k}_\perp^2-4(1-x)^2mM-4xM_v^2+(1-x)^2\bm{\Delta}_\perp^2)\cos(\overline{\chi}^{(v)}_b-\overline{\chi}^{(v)}_c)\sin(\bm{b}_\perp\cdot\bm{\Delta}_\perp+\delta\chi^{(v)}_b+\delta\chi^{(v)}_c)]\bigg\},\\
&\frac{k^j}{M}\rho^{j(v)}_{_\textrm{LT}}(x,\bm{b}_\perp,\bm{k}_\perp)\nonumber\\
=&\int\frac{d^2\bm{\Delta}_\perp}{(2\pi)^2}\frac{32g_v^2}{3\pi^3MD^2}(1-x)^2\bigg\{(m+xM)[4\bm{k}_\perp^2\cos\overline{\chi}^{(v)}_a\cos(\bm{b}_\perp\cdot\bm{\Delta}_\perp+\delta\chi^{(v)}_a)-2(1-x)\bm{k}_\perp\cdot\bm{\Delta}_\perp\sin\overline{\chi}^{(v)}_a\sin(\bm{b}_\perp\cdot\bm{\Delta}_\perp+\delta^{(v)}_a)]\nonumber\\
&\quad\quad +\frac{1}{M_v^2}(m+M)[4\bm{k}_\perp^2(4\bm{k}_\perp^2-4(1-x)^2mM-4xM_v^2+(1-x)^2\bm{\Delta}_\perp^2)\cos(\overline{\chi}^{(v)}_b-\overline{\chi}^{(v)}_c)\cos(\bm{b}_\perp\cdot\bm{\Delta}_\perp+\delta\chi^{(v)}_b+\delta\chi^{(v)}_c)\nonumber\\
&\quad\quad -2(1-x)\bm{k}_\perp\cdot\bm{\Delta}_\perp(4\bm{k}_\perp^2+4(1-x)^2mM+4xM_v^2-(1-x)^2\bm{\Delta}_\perp^2)\sin(\overline{\chi}^{(v)}_b-\overline{\chi}^{(v)}_c)\sin(\bm{b}_\perp\cdot\bm{\Delta}_\perp+\delta\chi^{(v)}_b+\delta\chi^{(v)}_c)\nonumber\\
&\quad\quad +8(1-x)(\bm{k}_\perp\cdot\bm{\Delta}_\perp)^2\cos(\overline{\chi}^{(v)}_b-\overline{\chi}^{(v)}_c)\cos(\bm{b}_\perp\cdot\bm{\Delta}_\perp+\delta\chi^{(v)}_b+\delta\chi^{(v)}_c)]\bigg\},\\
&\frac{k^j}{M}\rho^{j(v)}_{_\textrm{TU}}(x,\bm{b}_\perp,\bm{k}_\perp)\nonumber\\
=&\int\frac{d^2\bm{\Delta}_\perp}{(2\pi)^2}\frac{64g_v^2}{3\pi^3MD^2}(1-x)^3\bm{\Delta}_\perp\times\bm{k}_\perp\bigg\{-x(m+xM)\sin(\bm{b}_\perp\cdot\bm{\Delta}_\perp)\nonumber\\
&\quad\quad -\frac{1}{8M_v^2}(m+M)[4(1-x)\bm{k}_\perp\cdot\bm{\Delta}_\perp\sin(\overline{\chi}^{(v)}_b-\overline{\chi}^{(v)}_c)\cos(\bm{b}_\perp\cdot\bm{\Delta}_\perp+\delta\chi^{(v)}_b+\delta\chi^{(v)}_c)\nonumber\\
&\quad\quad +(4\bm{k}_\perp^2-4(1-x)^2mM-4xM_v^2+(1-x)^2\bm{\Delta}_\perp^2)\cos(\overline{\chi}^{(v)}_b-\overline{\chi}^{(v)}_c)\sin(\bm{b}_\perp\cdot\bm{\Delta}_\perp+\delta\chi^{(v)}_b+\delta\chi^{(v)}_c)]\bigg\},
\end{align}
\begin{align}
&\frac{k^j}{M}\rho^{j(v)}_{_\textrm{TL}}(x,\bm{b}_\perp,\bm{k}_\perp)\nonumber\\
=&\int\frac{d^2\bm{\Delta}_\perp}{(2\pi)^2}\frac{32g_v^2}{3\pi^3MD^2}(1-x)^2\bigg\{4x(m+xM)\bm{k}_\perp^2\cos(\bm{b}_\perp\cdot\bm{\Delta}_\perp)\nonumber\\
&\quad\quad +\frac{1}{M_v^2}(m+M)[4\bm{k}_\perp^2(4\bm{k}_\perp^2-4(1-x)^2mM-4xM_v^2+(1-x)^2\bm{\Delta}_\perp^2)\cos(\overline{\chi}^{(v)}_b-\overline{\chi}^{(v)}_c)\cos(\bm{b}_\perp\cdot\bm{\Delta}_\perp+\delta\chi^{(v)}_b+\delta\chi^{(v)}_c)\nonumber\\
&\quad\quad -2(1-x)\bm{k}_\perp\cdot\bm{\Delta}_\perp(4\bm{k}_\perp^2+4(1-x)^2mM+4xM_v^2-(1-x)^2\bm{\Delta}_\perp^2)\sin(\overline{\chi}^{(v)}_b-\overline{\chi}^{(v)}_c)\sin(\bm{b}_\perp\cdot\bm{\Delta}_\perp+\delta\chi^{(v)}_b+\delta\chi^{(v)}_c)\nonumber\\
&\quad\quad +8(1-x)(\bm{k}_\perp\cdot\bm{\Delta}_\perp)^2\cos(\overline{\chi}^{(v)}_b-\overline{\chi}^{(v)}_c)\cos(\bm{b}_\perp\cdot\bm{\Delta}_\perp+\delta\chi^{(v)}_b+\delta\chi^{(v)}_c)]\bigg\},\\
&\rho^{(v)}_{_\textrm{TT}}(x,\bm{b}_\perp,\bm{k}_\perp)\nonumber\\
=&\int\frac{d^2\bm{\Delta}_\perp}{(2\pi)^2}\frac{32g_v^2}{3\pi^3D^2}(1-x)\bigg\{2x[(1-x)^2\bm{\Delta}_\perp^2-4\bm{k}_\perp^2]\cos\overline{\chi}^{(v)}_a\cos(\bm{b}_\perp\cdot\bm{\Delta}_\perp+\delta\chi^{(v)}_a)\nonumber\\
&\quad\quad -\frac{1}{8M_v^2}[(4\bm{k}_\perp^2-4(1-x)^2mM-4xM_v^2+(1-x)^2\bm{\Delta}_\perp^2)^2-16(1-x)^2(\bm{k}_\perp\cdot\bm{\Delta}_\perp)^2]\cos(\bm{b}_\perp\cdot\bm{\Delta}_\perp+2\delta\chi^{(v)}_b)\bigg\},\\
&\rho^{\perp(v)}_{_\textrm{TT}}(x,\bm{b}_\perp,\bm{k}_\perp)\nonumber\\
=&\int\frac{d^2\bm{\Delta}_\perp}{(2\pi)^2}\frac{256g_v^2}{3\pi^3}\frac{x(1-x)^2\bm{k}_\perp\times\bm{\Delta}_\perp\sin\overline{\chi}^{(v)}_a\sin(\bm{b}_\perp\cdot\bm{\Delta}_\perp+\delta\chi^{(v)}_a)}{[(4\bm{k}_\perp^2+4L_v^2+(1-x)^2\bm{\Delta}_\perp^2)^2-16(1-x)^2(\bm{k}_\perp\cdot\bm{\Delta}_\perp)^2]^2},\label{pretv}
\end{align}
where the denominator
\begin{equation}
D=[4\bm{k}_\perp^2+4L_v^2+(1-x)^2\bm{\Delta}_\perp^2]^2-16(1-x)^2(\bm{k}_\perp\cdot\bm{\Delta}_\perp)^2.
\end{equation}
As seen from the expressions Eqs. (\ref{uuv})-(\ref{pretv}), the relations found in the scalar spectator case (\ref{rel1})-(\ref{rel4}) are violated in the axial-vector case. If only the transverse polariations of the axial-vector spectator are taken into account, we will still have a relation between unpol-longitudinal and longi-unpolarized distributions but with an opposite sign to the one with scalar spectator:
\begin{equation}
\rho^{(v)}_{_\textrm{UL}}(x,\bm{b}_\perp,\bm{k}_\perp)=\rho^{(v)}_{_\textrm{LU}}(x,\bm{b}_\perp,\bm{k}_\perp).
\end{equation}

With the longi-unpolarized and unpol-longitudinal Wigner distributions, one can define the $x$ dependent orbital angular momentum $\ell_z(x)$ and spin-orbit correlator $\mathcal{C}_z(x)$ as
\begin{eqnarray}
\ell_z(x)&=&\int d^2\bm{b}_\perp d^2{k}_\perp \bm{b}_\perp\times\bm{k}_\perp\rho_{_\textrm{LU}}(x,\bm{b}_\perp,\bm{k}_\perp),\label{oam}\\
\mathcal{C}_z(x)&=&\int d^2\bm{b}_\perp d^2{k}_\perp \bm{b}_\perp\times\bm{k}_\perp\rho_{_\textrm{UL}}(x,\bm{b}_\perp,\bm{k}_\perp).\label{sl}
\end{eqnarray}
Substituting the corresponding distributions with Eqs. (\ref{uls}), (\ref{lus}), (\ref{ulv}) and (\ref{luv}), we get the expressions of the orbital angular momentum and spin-orbit correlator as
\begin{eqnarray}
\ell_z^{(s)}(x)&=&\frac{g_s^2(1-x)^4}{48\pi^2[xM_s^2+(1-x)m^2-x(1-x)M^2]^2},\\
\ell_z^{(v)}(x)&=&-\frac{g_v^2(1-x)^2[(1-x^2)-\frac{1}{2M_v^2}(1-x)^2(m+M)^2]}{72\pi^2[xM_v^2+(1-x)m^2-x(1-x)M^2]^2},\\
\mathcal{C}_z^{(s)}(x)&=&-\frac{g_s^2(1-x)^4}{48\pi^2[xM_s^2+(1-x)m^2-x(1-x)M^2]^2},\\
\mathcal{C}_z^{(v)}(x)&=&-\frac{g_v^2(1-x)^2[(1-x^2)+\frac{1}{2M_v^2}(1-x)^2(m+M)^2]}{72\pi^2[xM_v^2+(1-x)m^2-x(1-x)M^2]^2},
\end{eqnarray}

\section{Numerical results}

In this section, we provide the numerical results of quark mixing distributions which reflect quark orbital motions in the proton. For simplicity we only present the mixing distributions without transverse polarizations, and with the parameters and expressions one may easily get the numerical results with transverse polarizations if needed.

\begin{figure}
\includegraphics[width=0.35\textwidth]{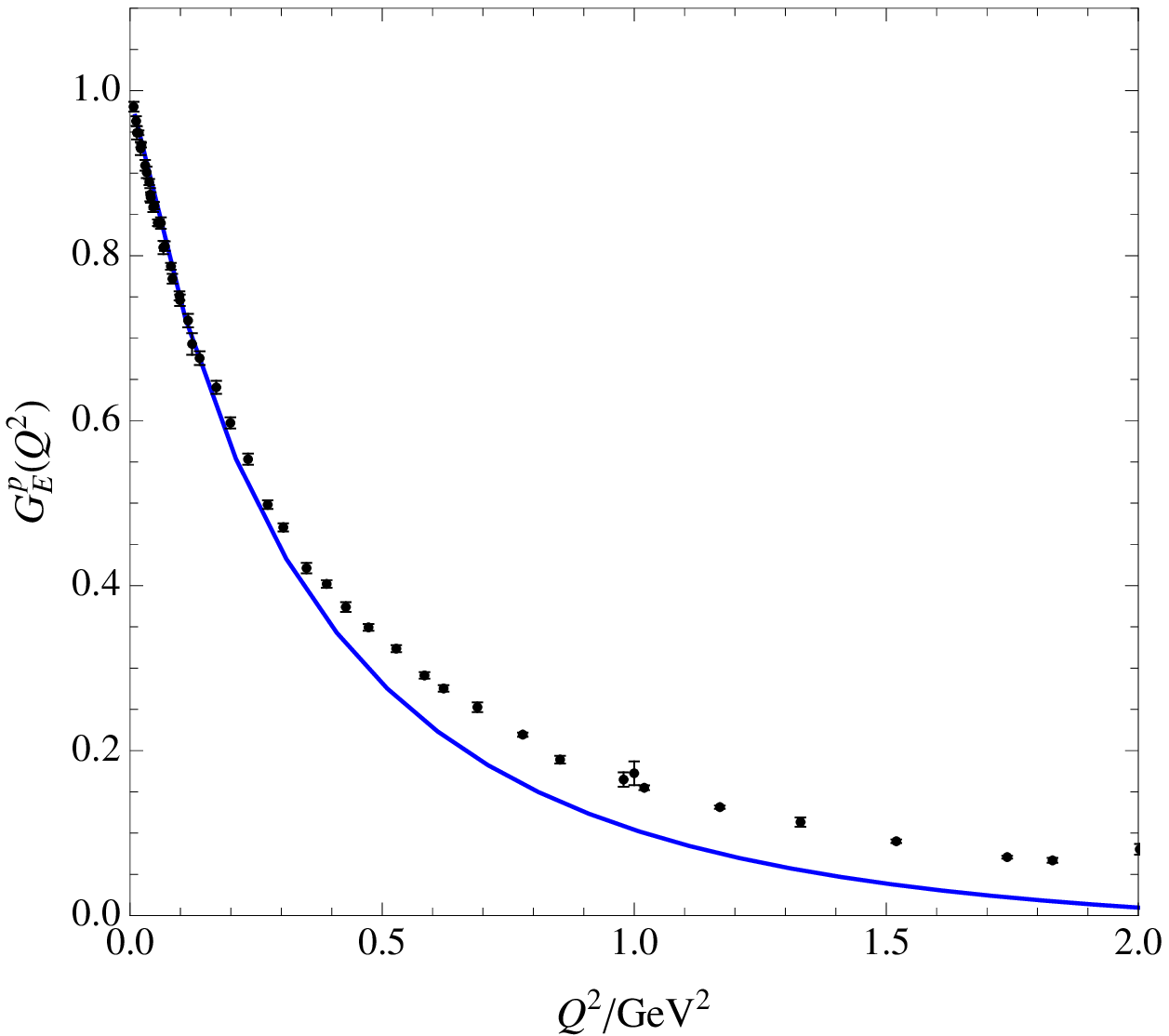}
\includegraphics[width=0.35\textwidth]{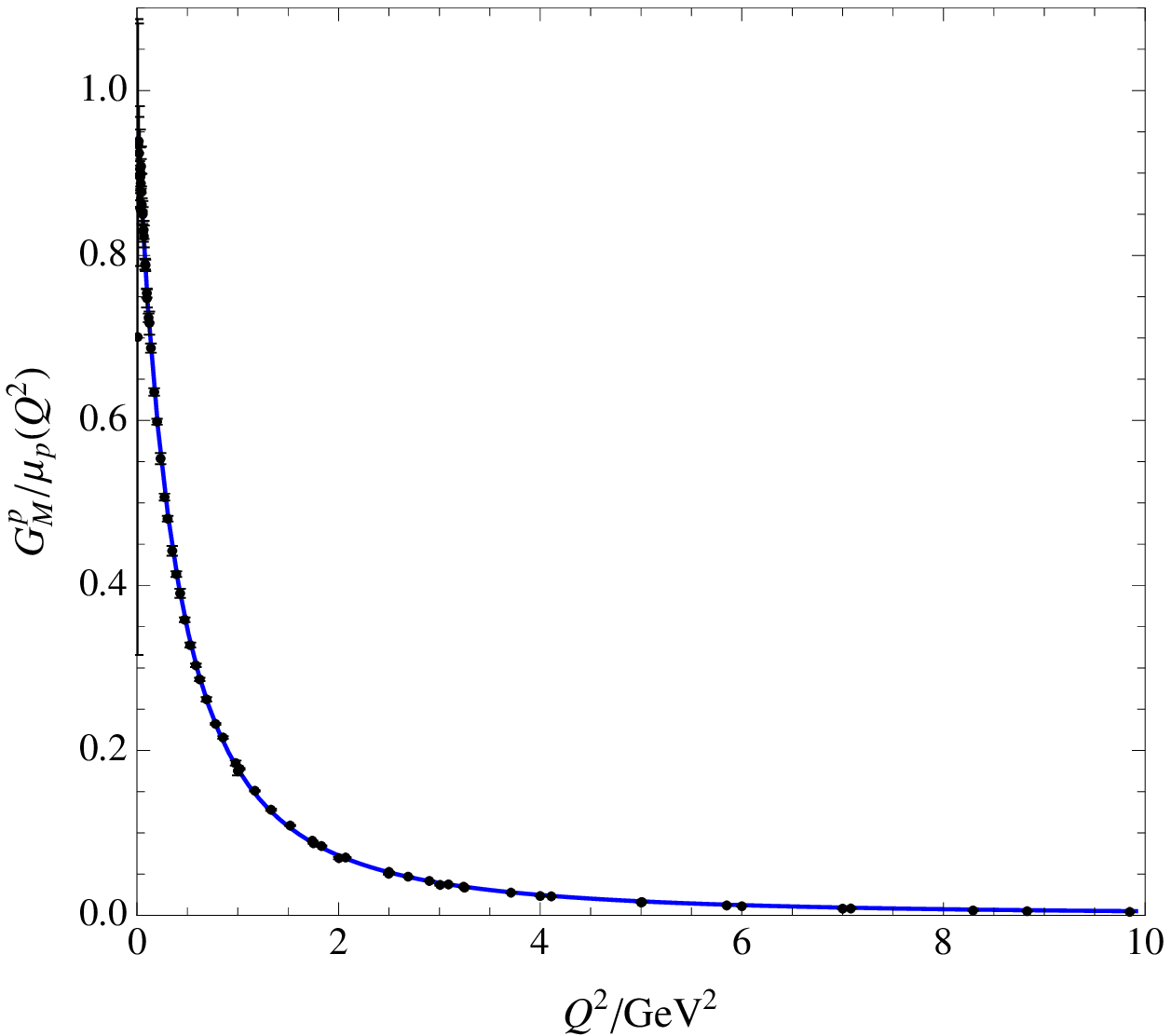}
\includegraphics[width=0.35\textwidth]{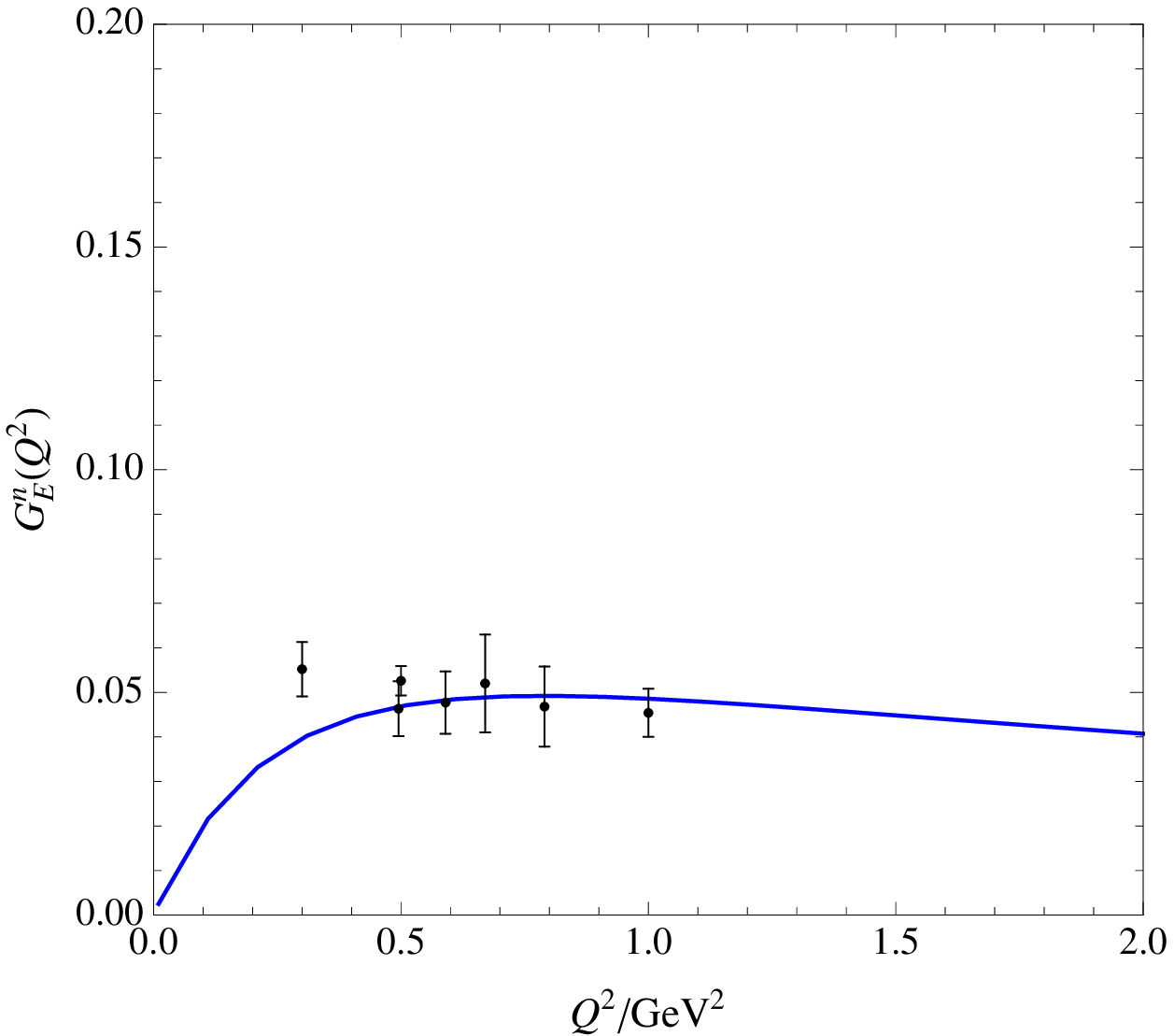}
\includegraphics[width=0.35\textwidth]{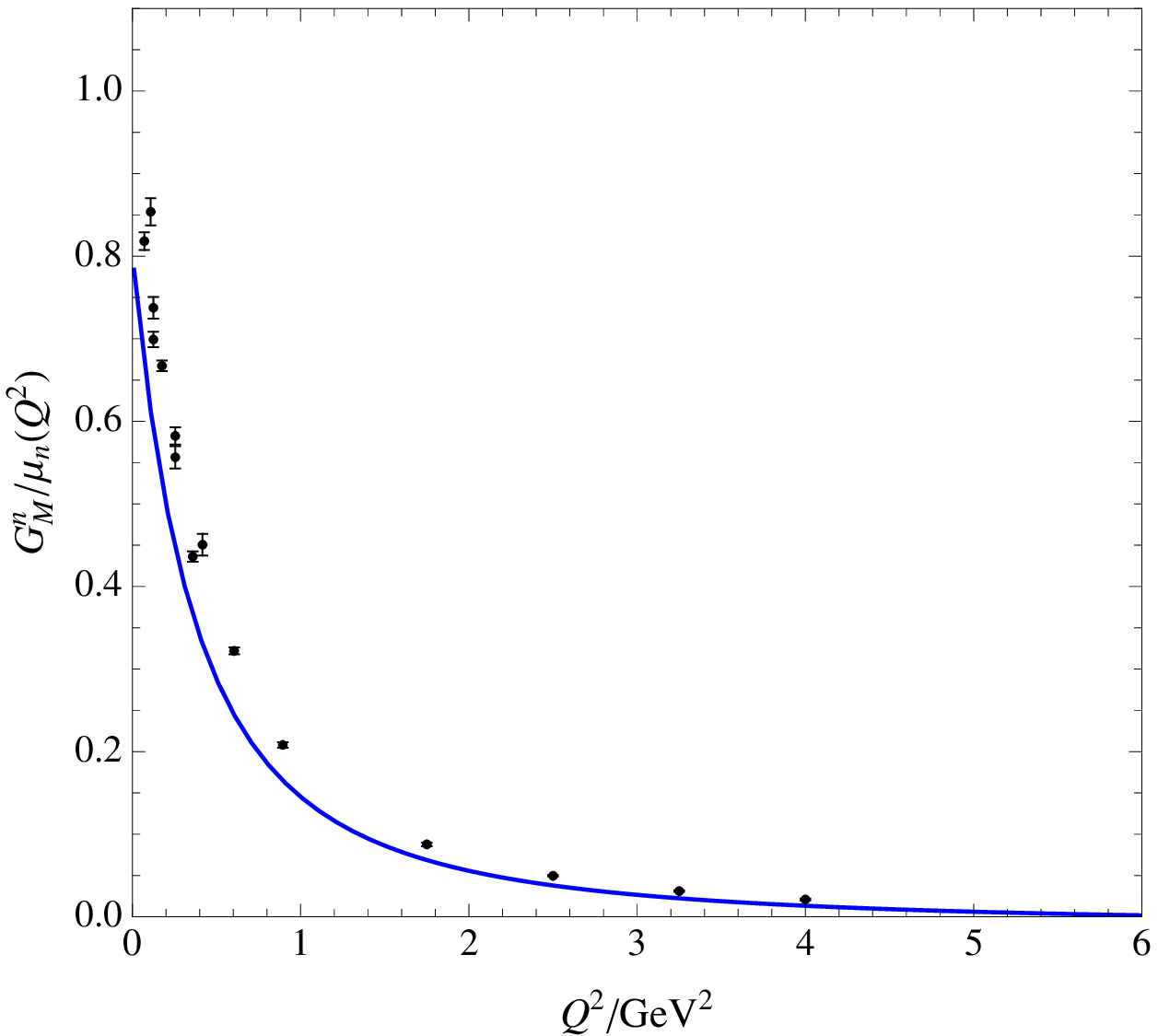}
\caption{(color online). The form factors of the proton (upper panels) and the neutron (lower panels). The curves are calculated from the model with parameters $m=451.4\,\textrm{MeV}$ and $M_d=705.3\,\textrm{MeV}$. The data are taken from Refs.~\cite{Arrington:2007ux,Walker:1993vj,Borkowski:1974mb,Kubon:2001rj,Glazier:2004ny,Markowitz:1993hx,Bruins:1995ns,Lung:1992bu,Zhu:2001md,Warren:2003ma,Rohe:1999sh}.\label{ff}}
\end{figure}
\begin{figure}
\includegraphics[width=0.3\textwidth]{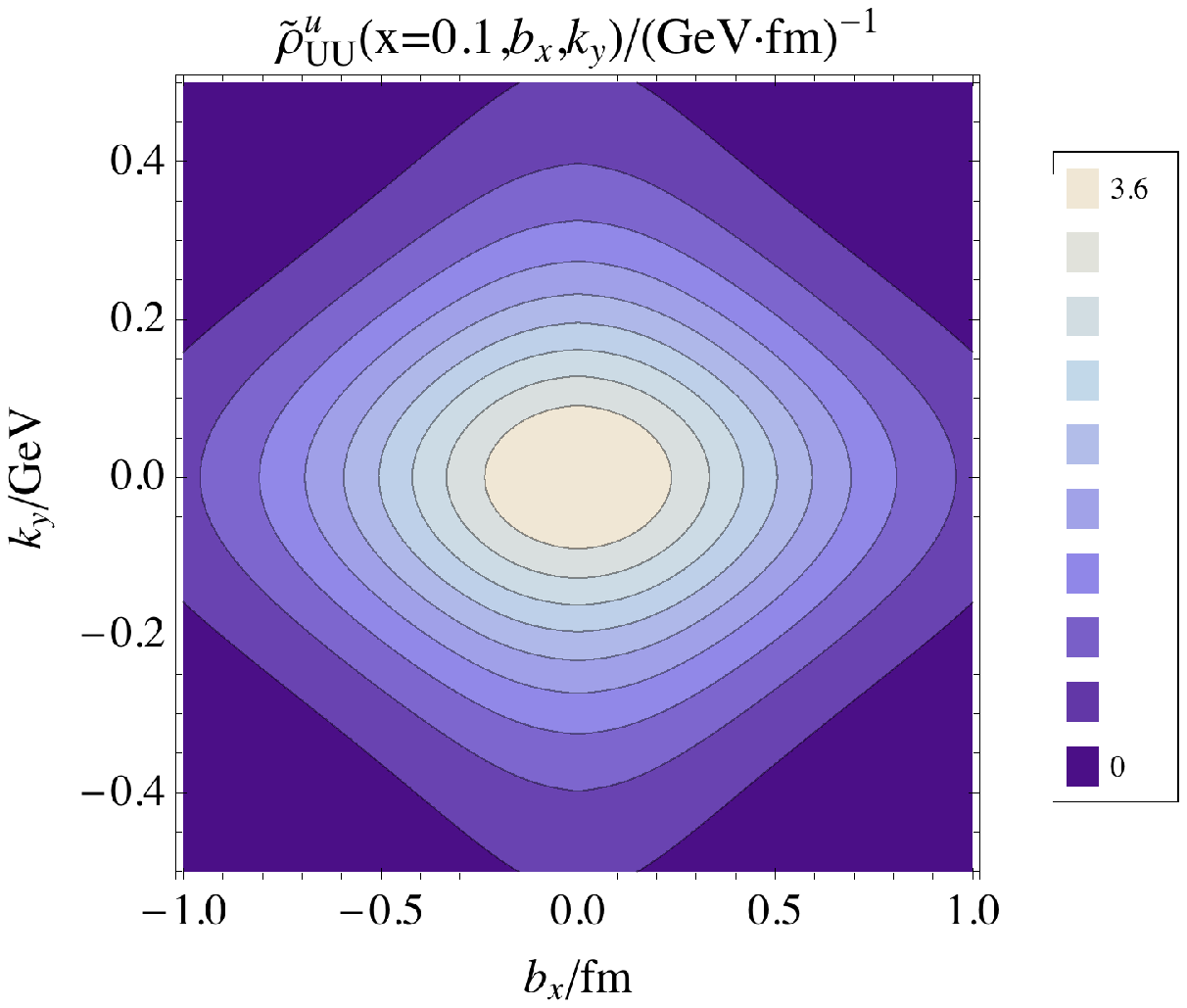}
\includegraphics[width=0.3\textwidth]{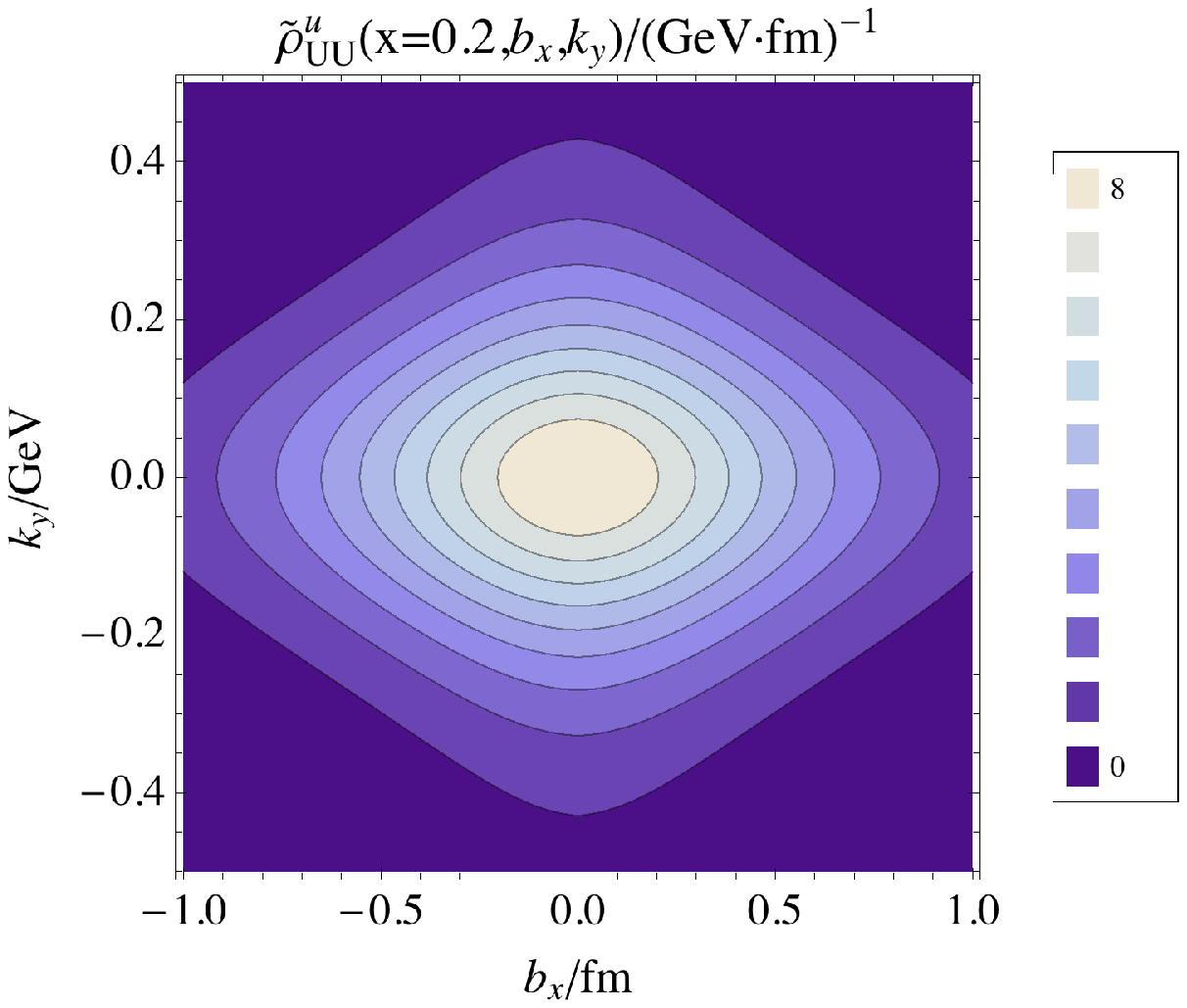}
\includegraphics[width=0.3\textwidth]{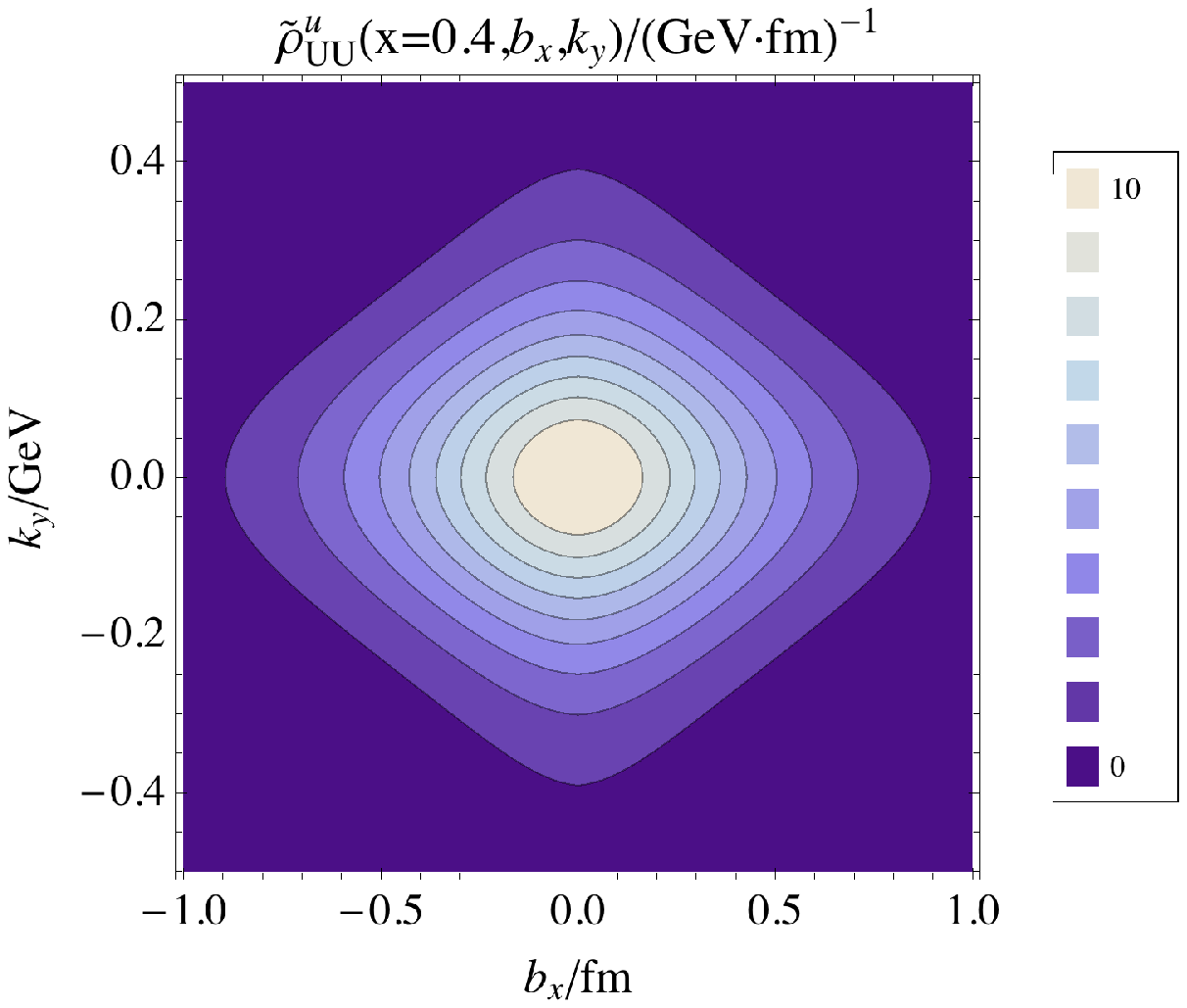}
\includegraphics[width=0.3\textwidth]{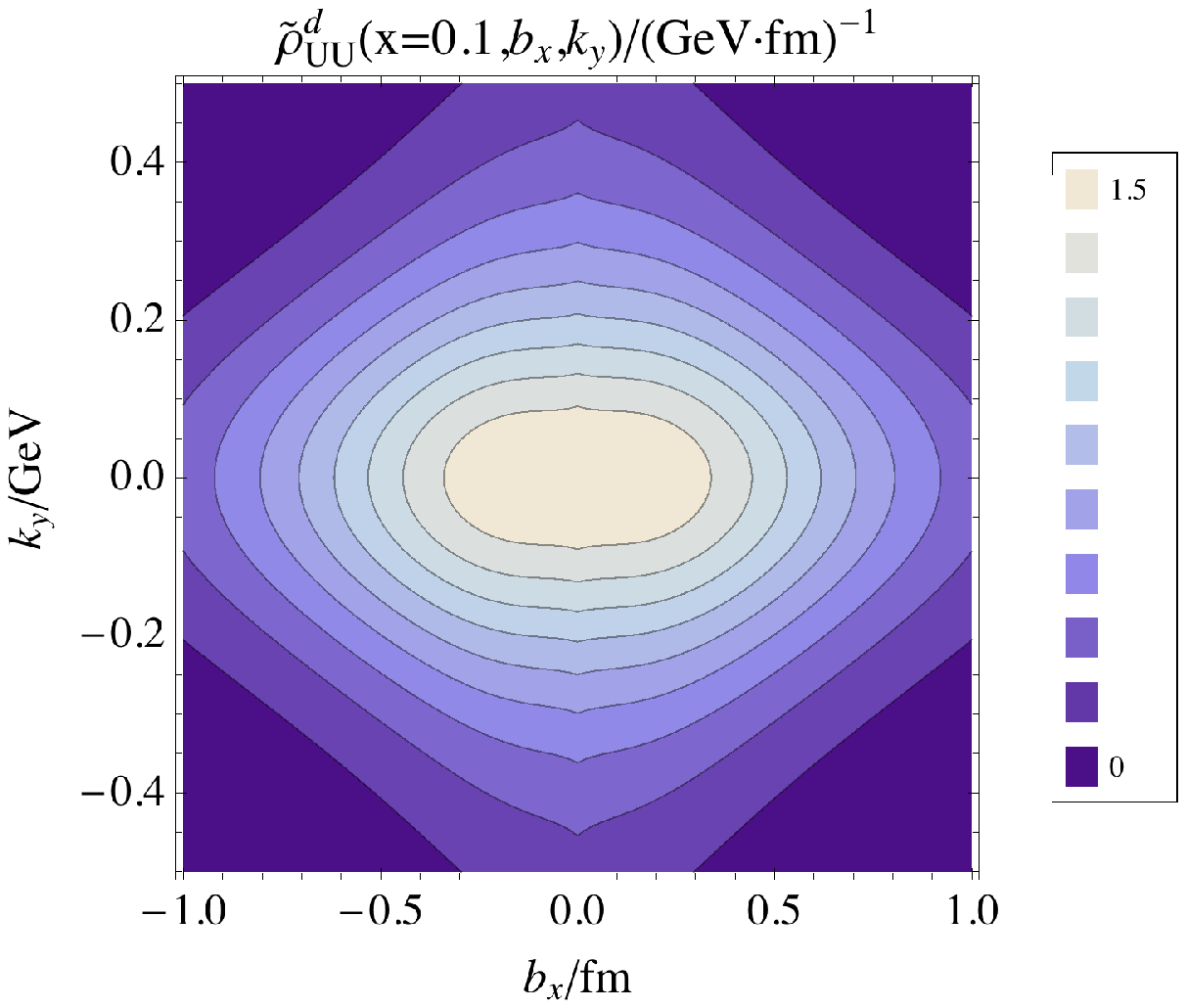}
\includegraphics[width=0.3\textwidth]{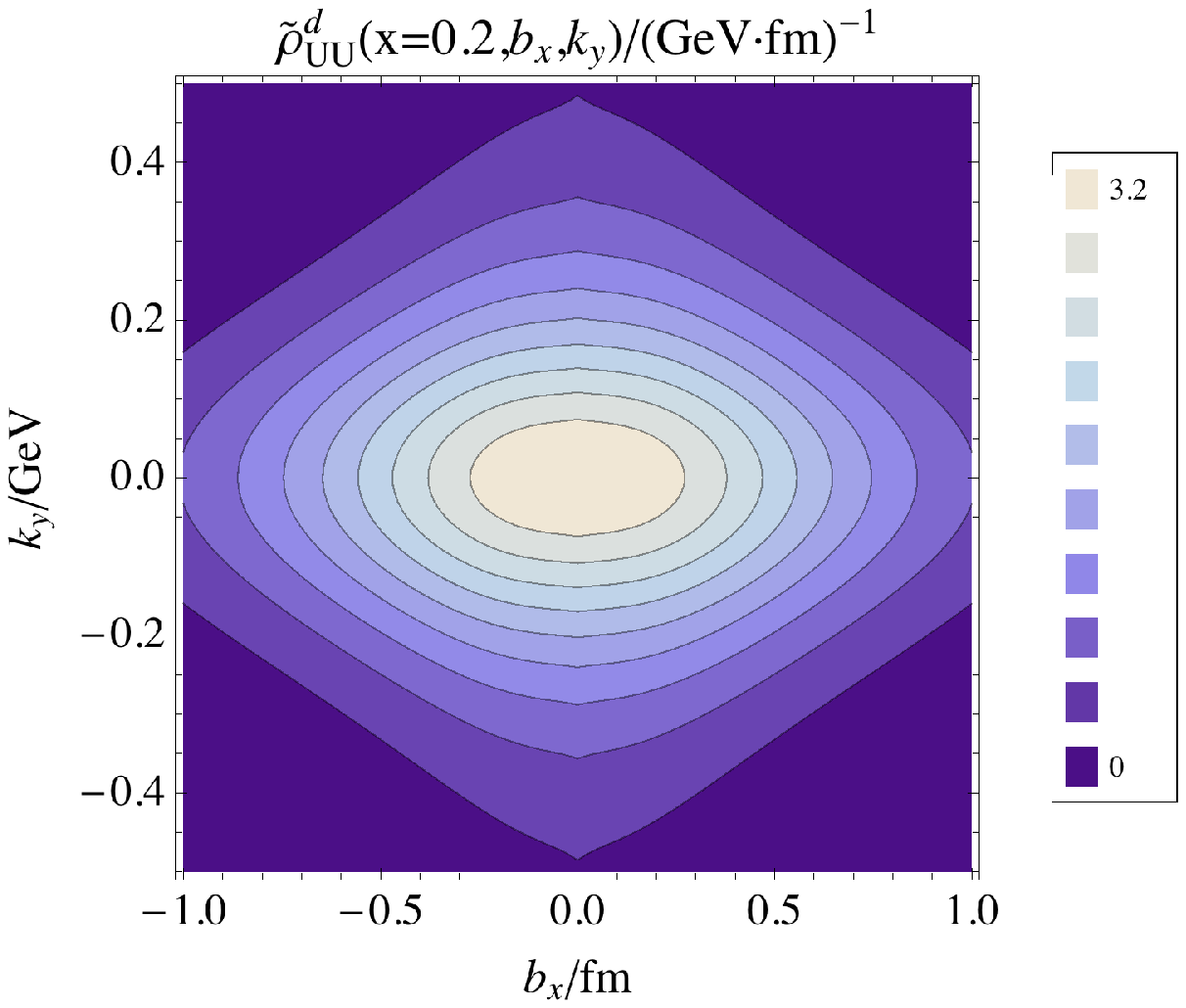}
\includegraphics[width=0.3\textwidth]{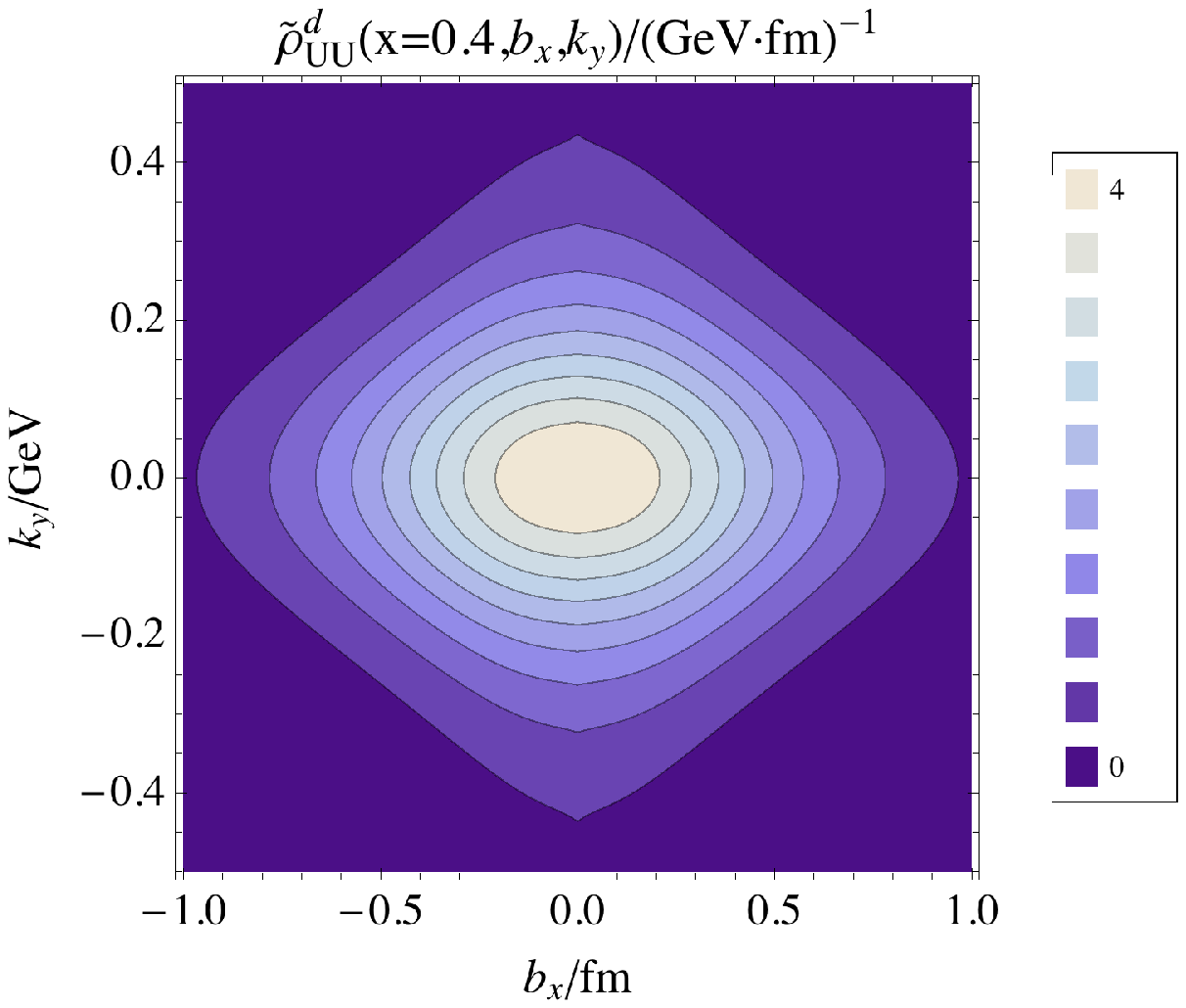}
\caption{(color online). Unpolarized mixing distributions $\tilde{\rho}_{_\textrm{UU}}(x,b_x,k_y)$ for $u$ quark (upper panels) and $d$ quark (lower panels) at $x=0.1$ (left column), $x=0.2$ (middle column) and $x=0.4$ (right column).\label{uumix}}
\includegraphics[width=0.3\textwidth]{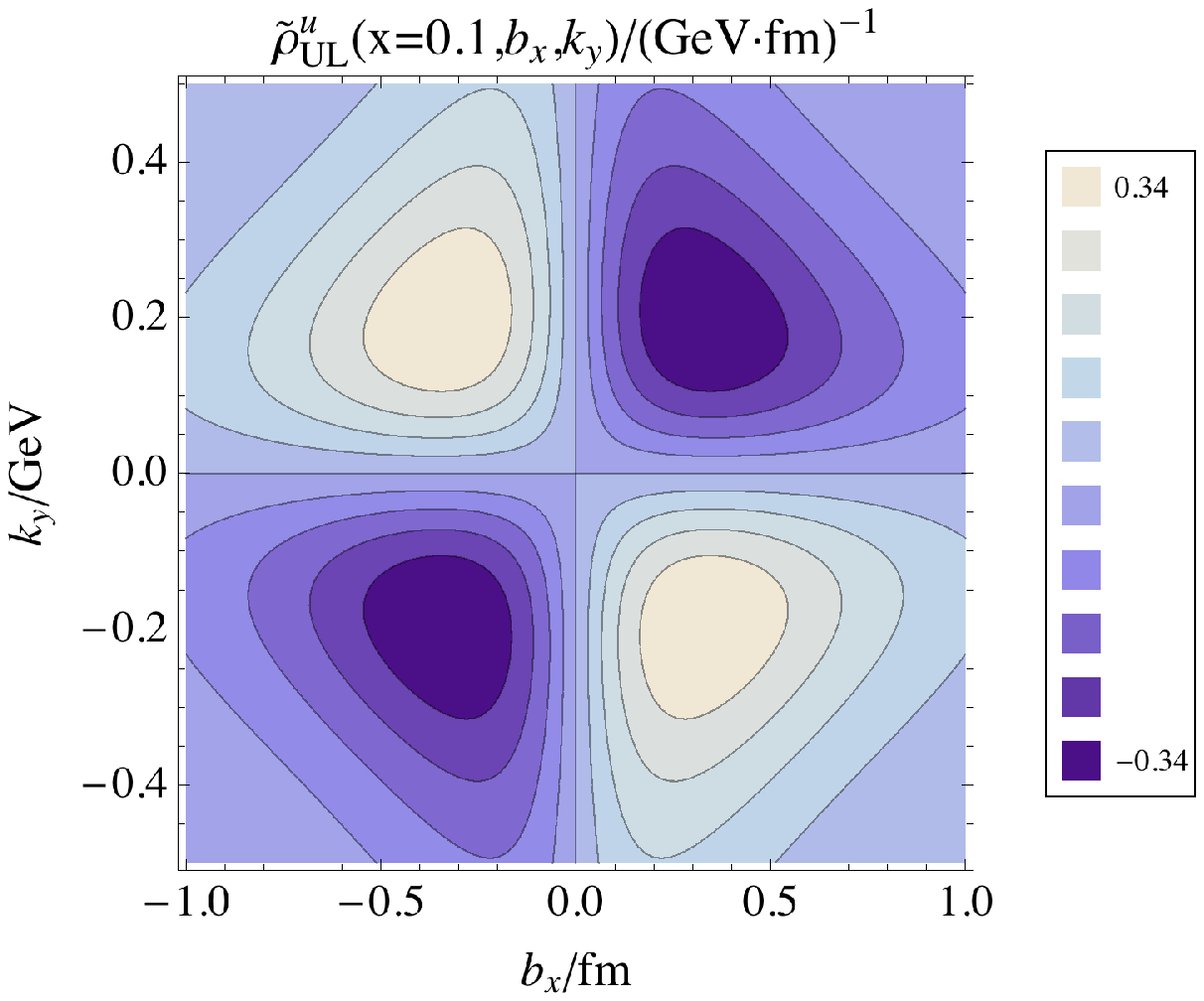}
\includegraphics[width=0.3\textwidth]{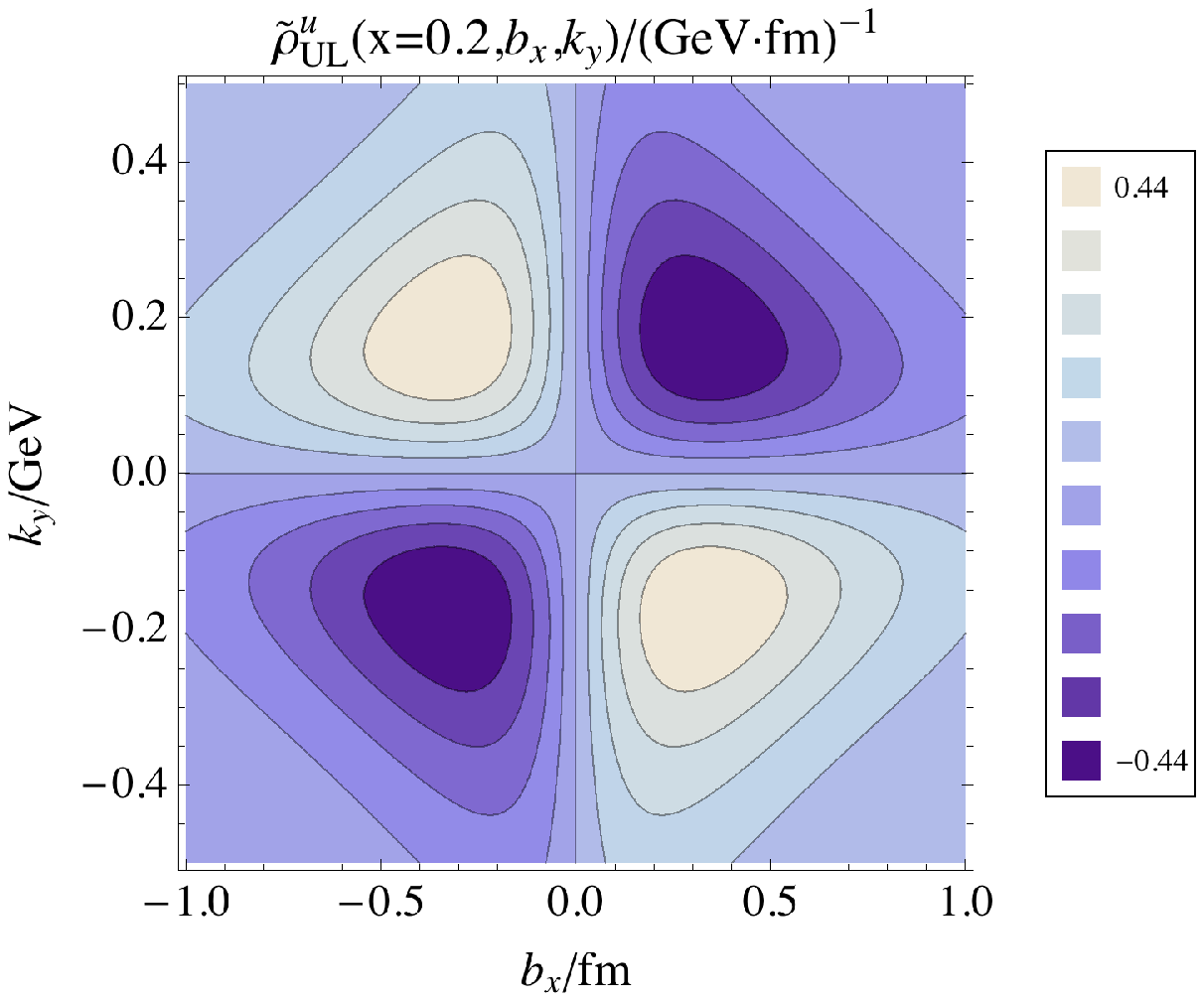}
\includegraphics[width=0.3\textwidth]{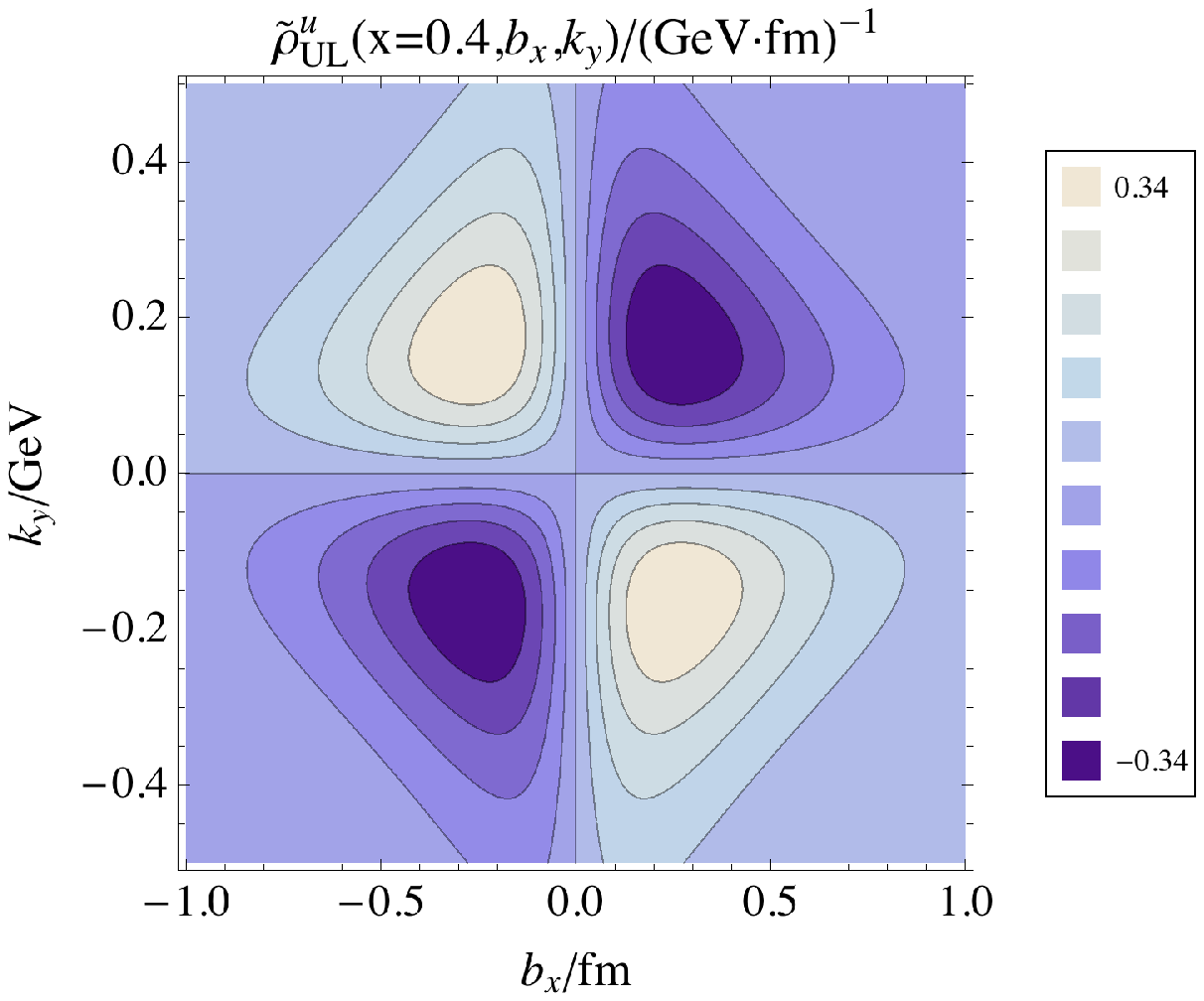}
\includegraphics[width=0.3\textwidth]{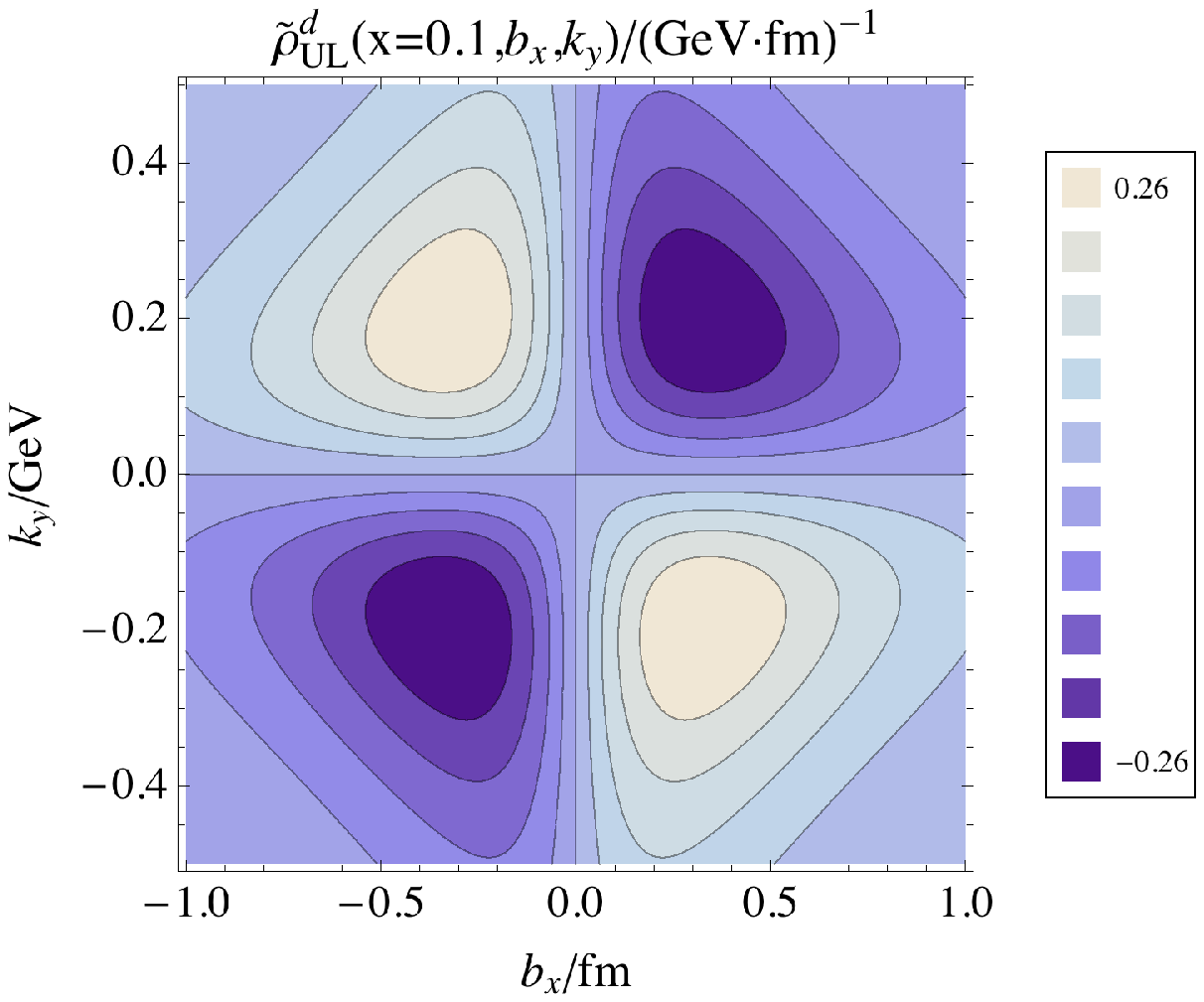}
\includegraphics[width=0.3\textwidth]{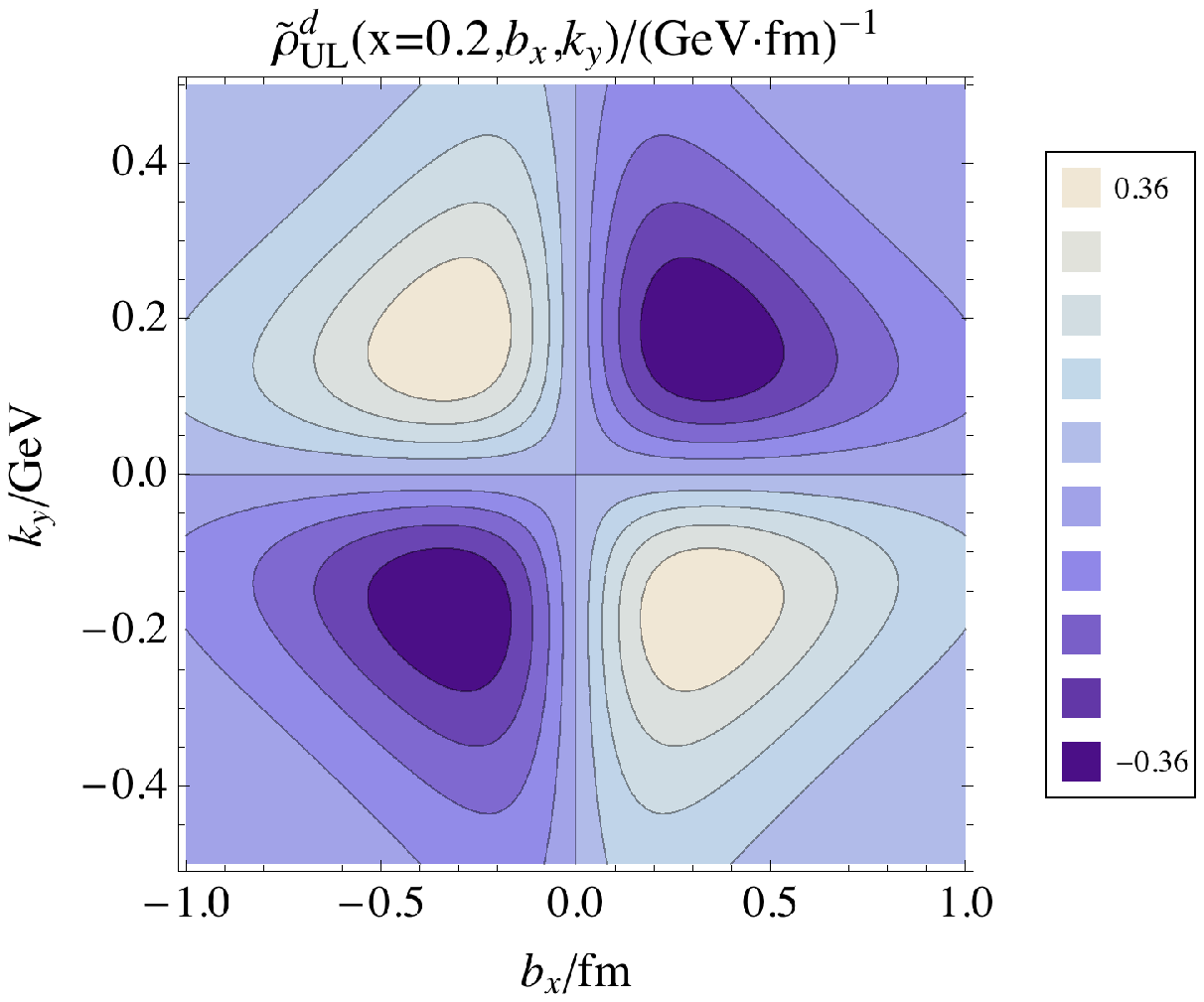}
\includegraphics[width=0.3\textwidth]{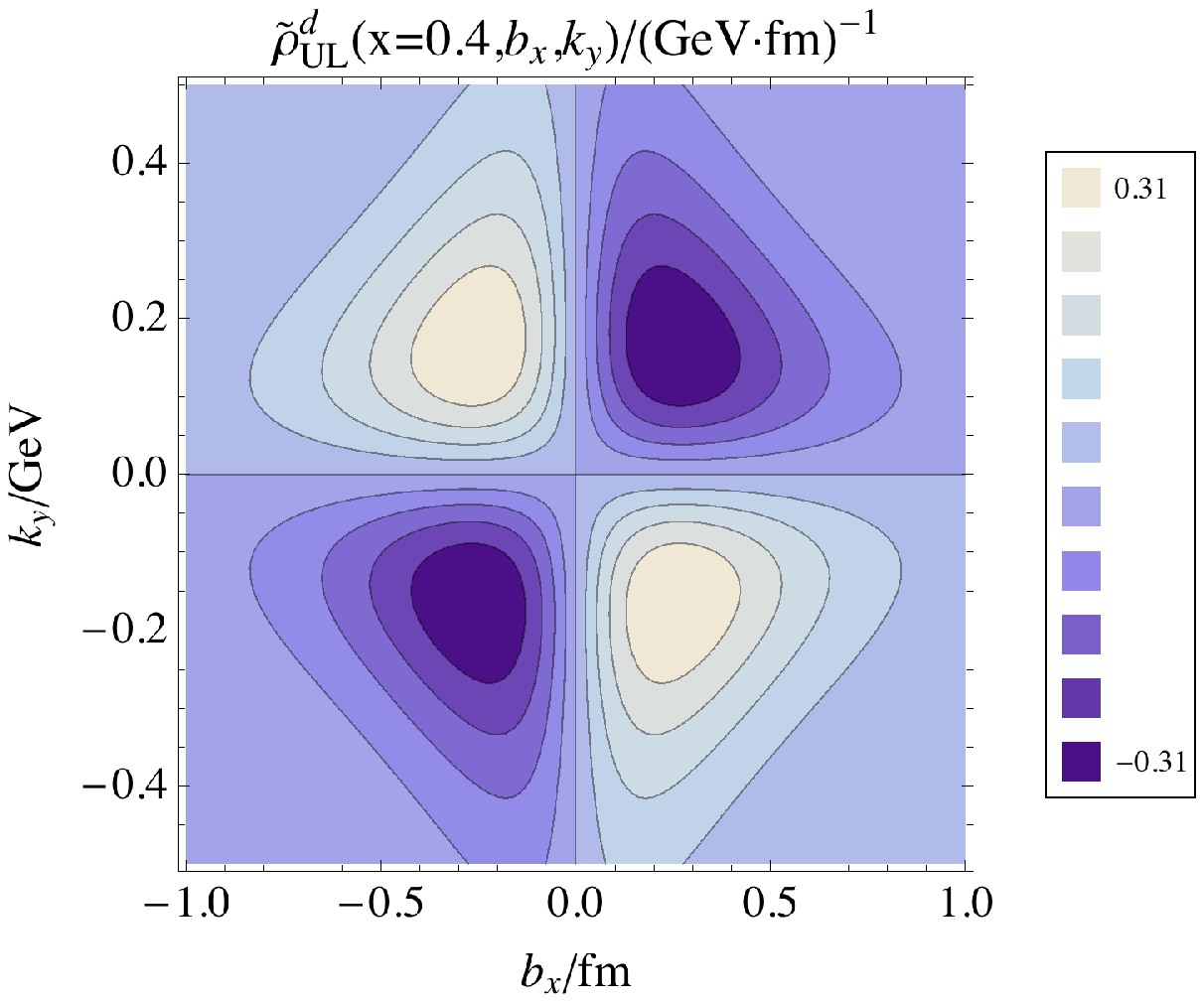}
\caption{(color online). Unpol-longitudinal mixing distributions $\tilde{\rho}_{_\textrm{UL}}(x,b_x,k_y)$ for $u$ quark (upper panels) and $d$ quark (lower panels) at $x=0.1$ (left column), $x=0.2$ (middle column) and $x=0.4$ (right column).\label{ulmix}}
\end{figure}
\begin{figure}
\includegraphics[width=0.3\textwidth]{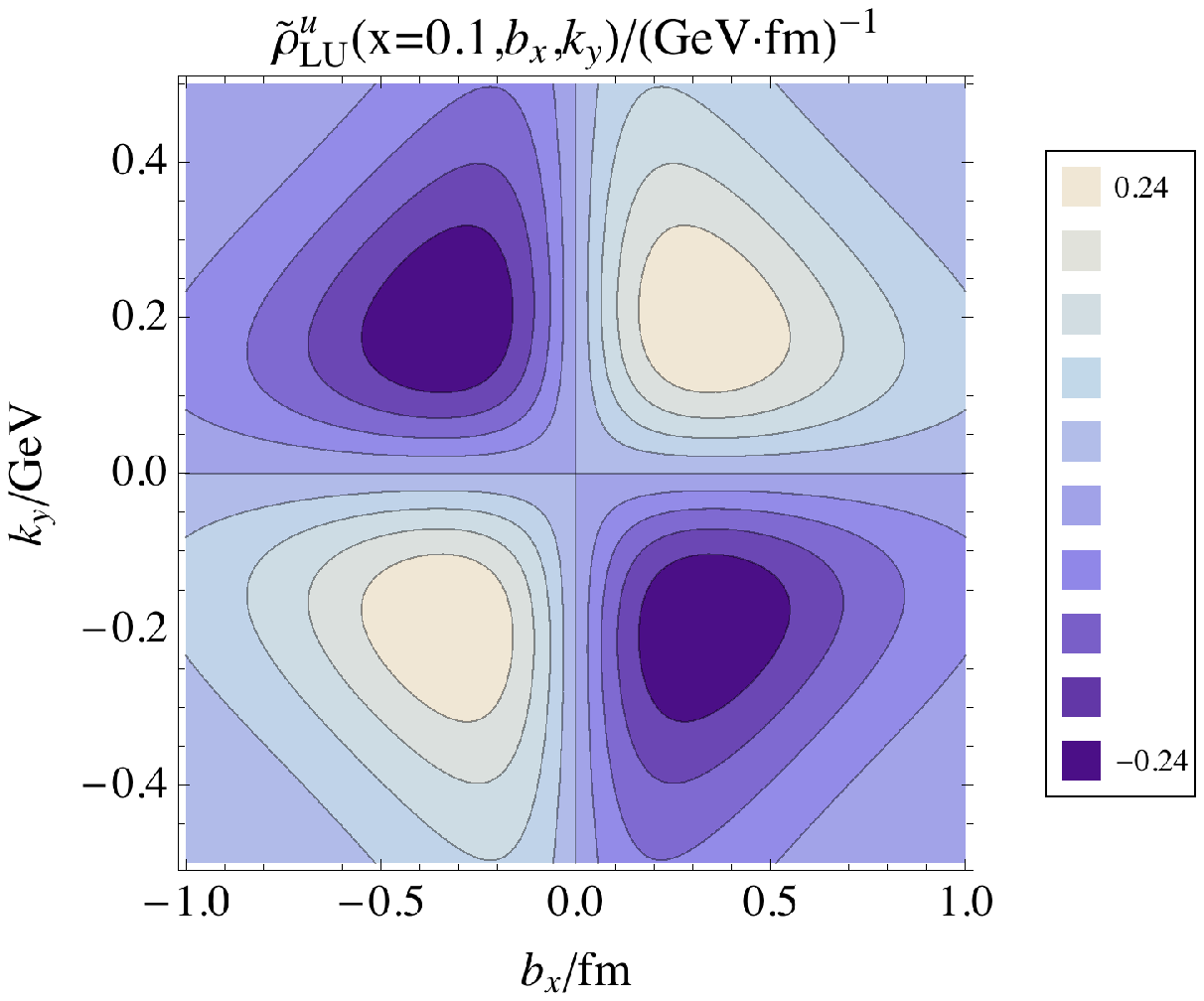}
\includegraphics[width=0.3\textwidth]{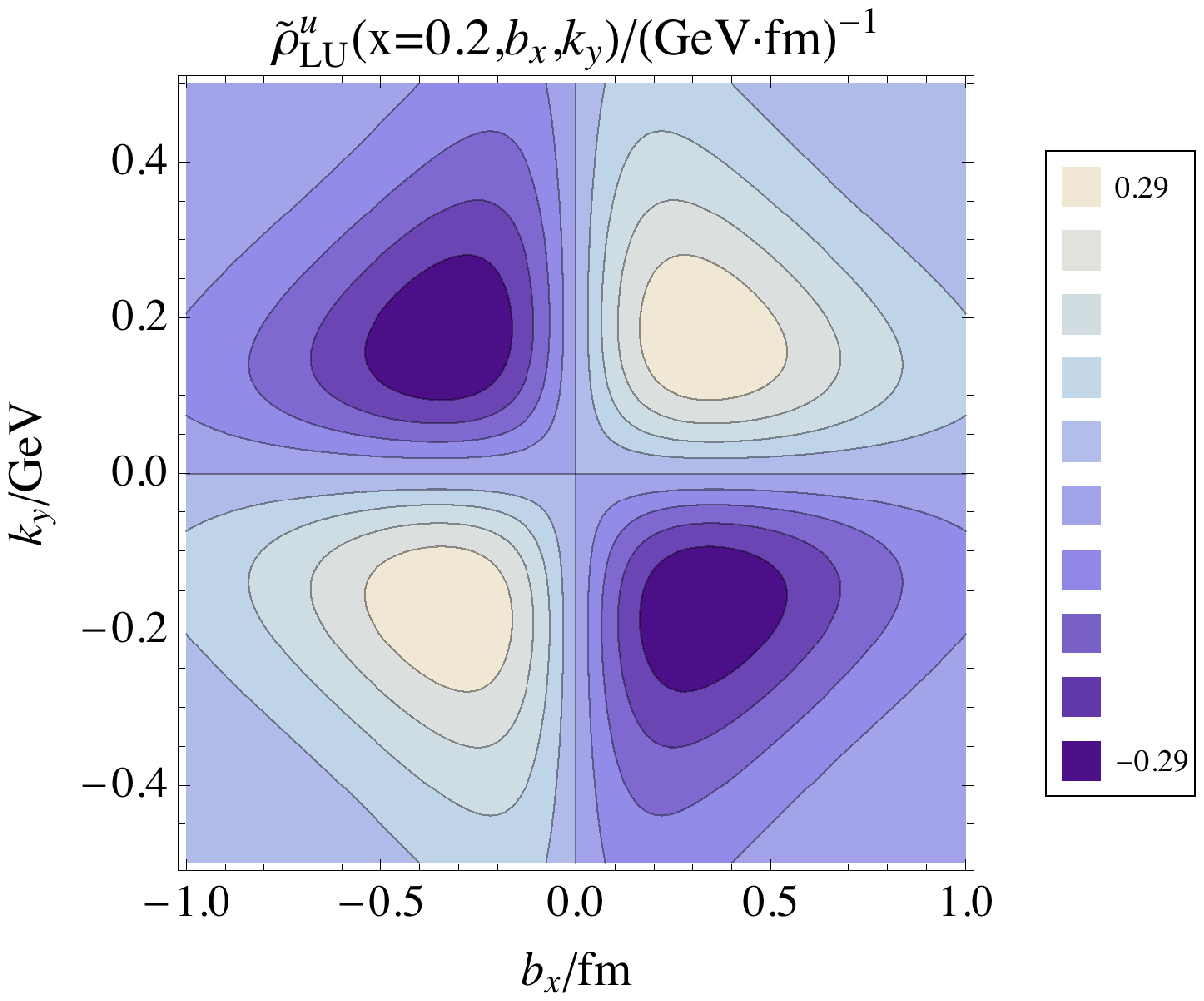}
\includegraphics[width=0.3\textwidth]{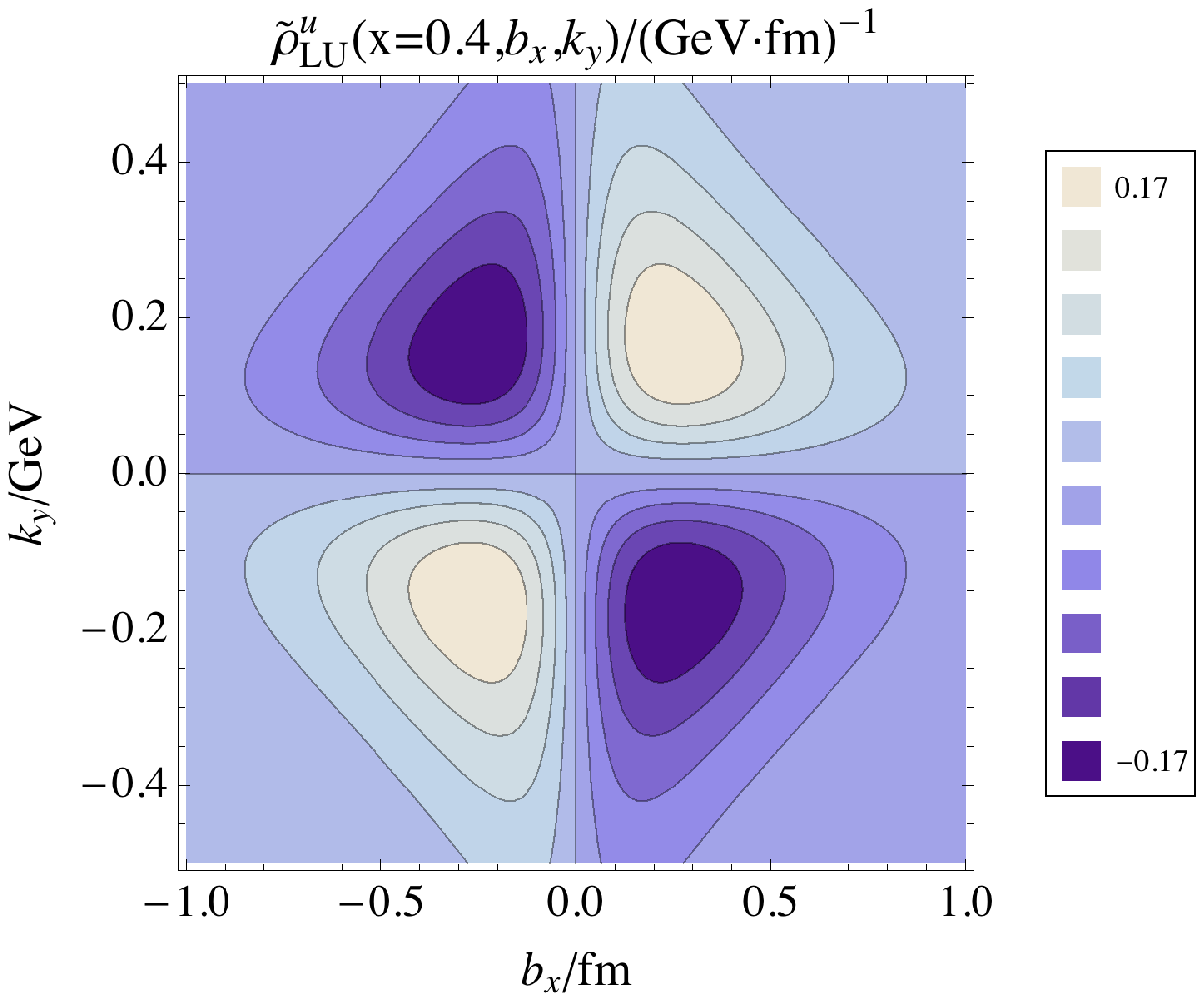}
\includegraphics[width=0.3\textwidth]{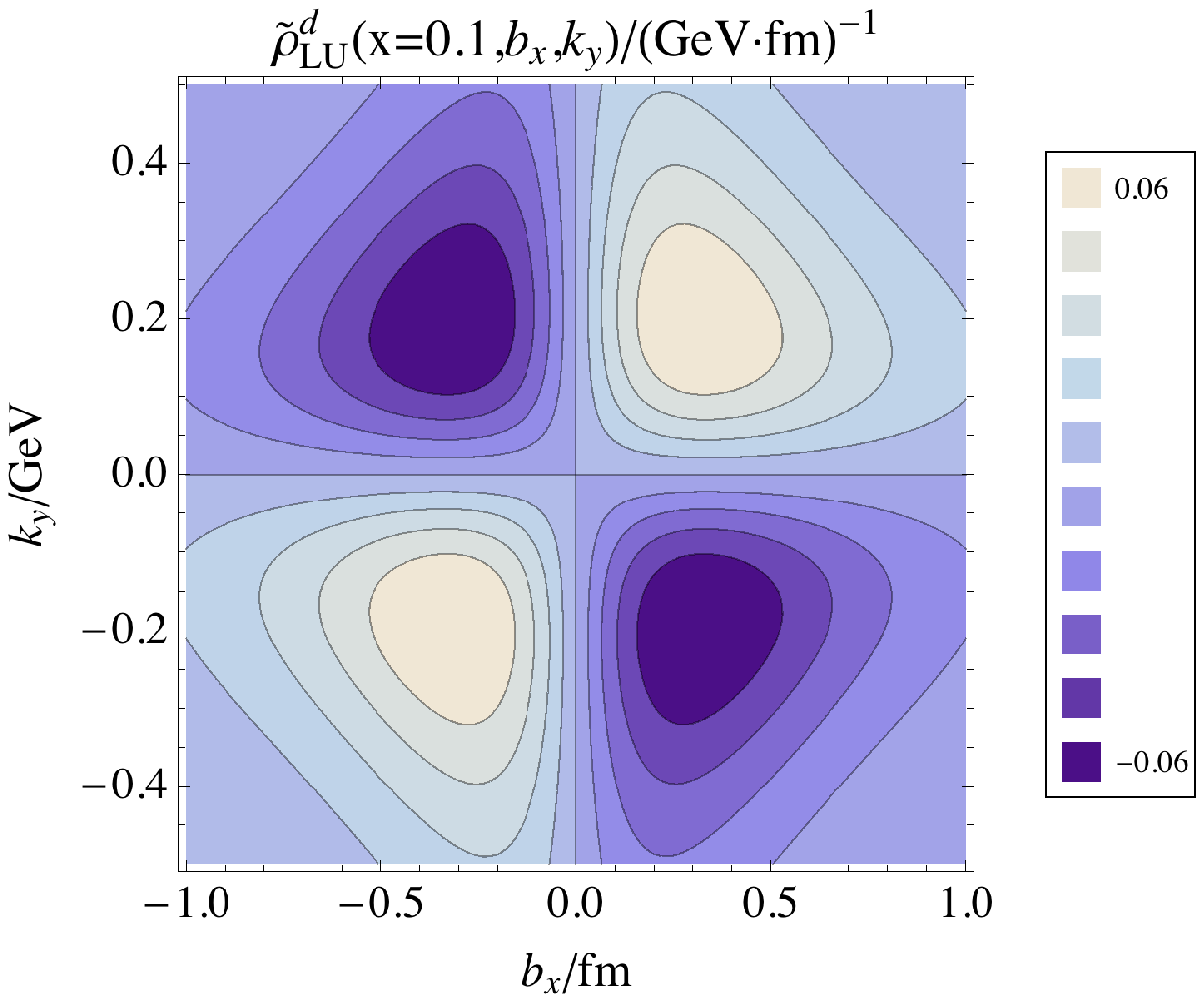}
\includegraphics[width=0.3\textwidth]{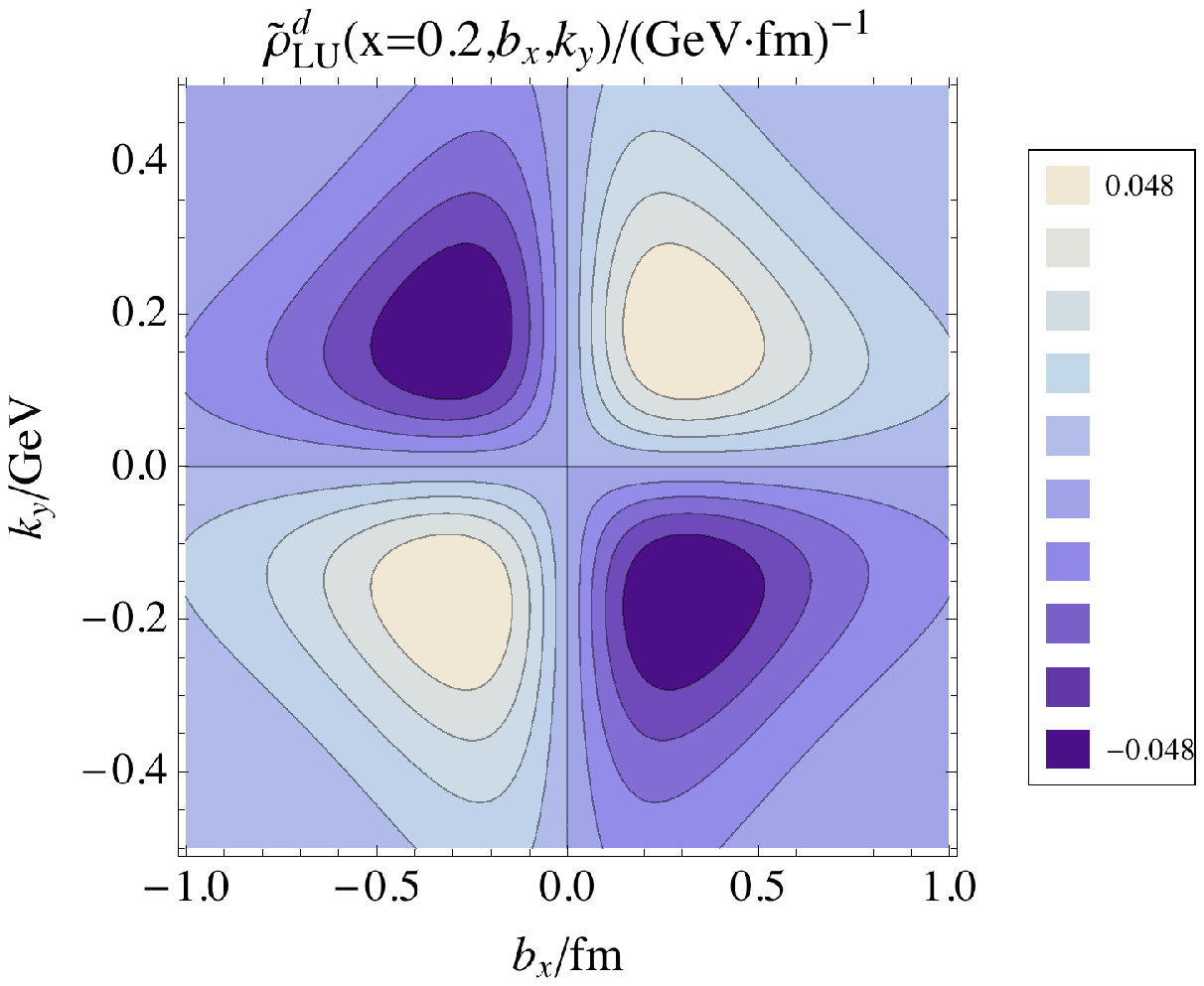}
\includegraphics[width=0.3\textwidth]{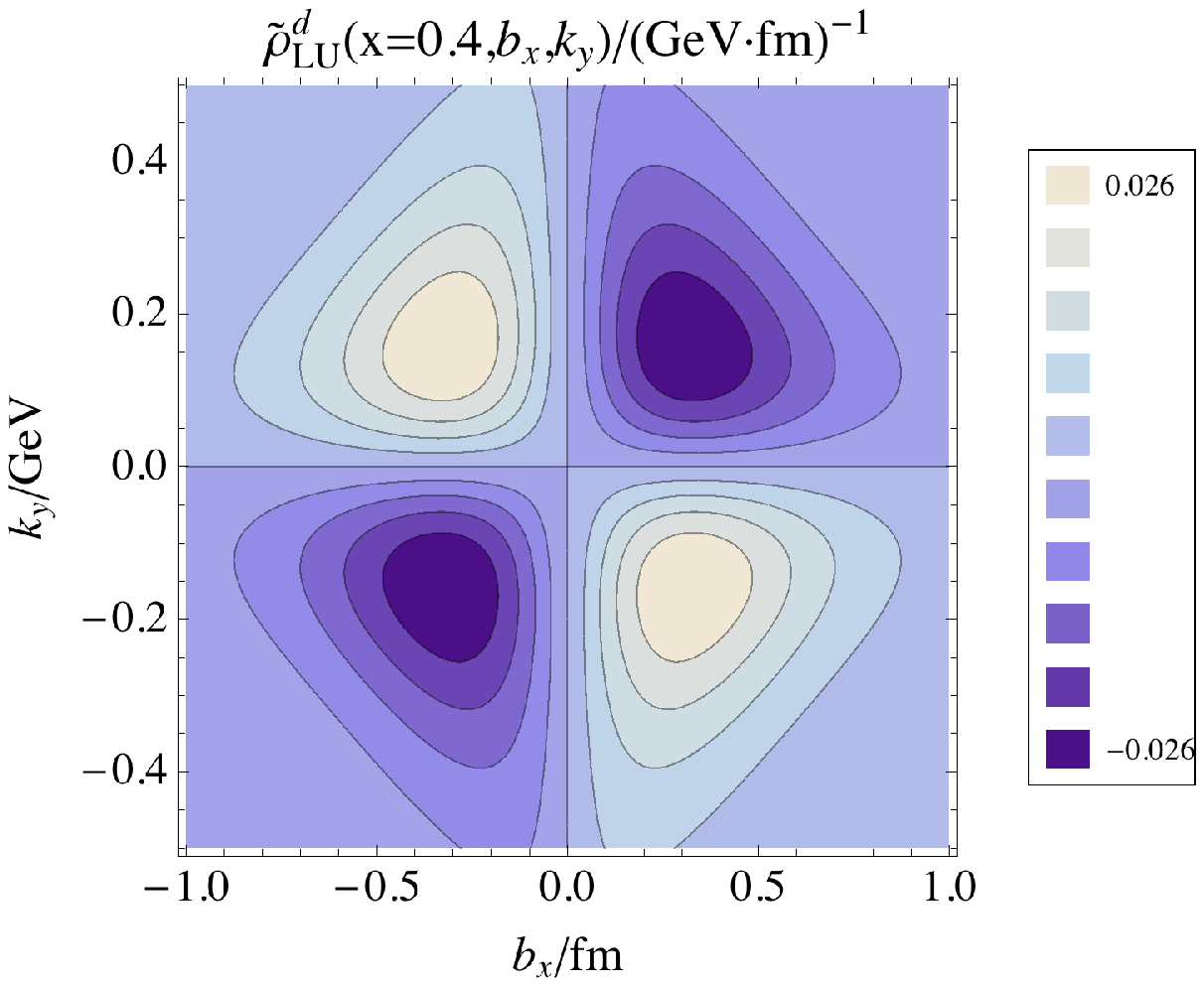}
\caption{(color online). Longi-unpolarized mixing distributions $\tilde{\rho}_{_\textrm{LU}}(x,b_x,k_y)$ for $u$ quark (upper panels) and $d$ quark (lower panels) at $x=0.1$ (left column), $x=0.2$ (middle column) and $x=0.4$ (right column).\label{lumix}}
\includegraphics[width=0.3\textwidth]{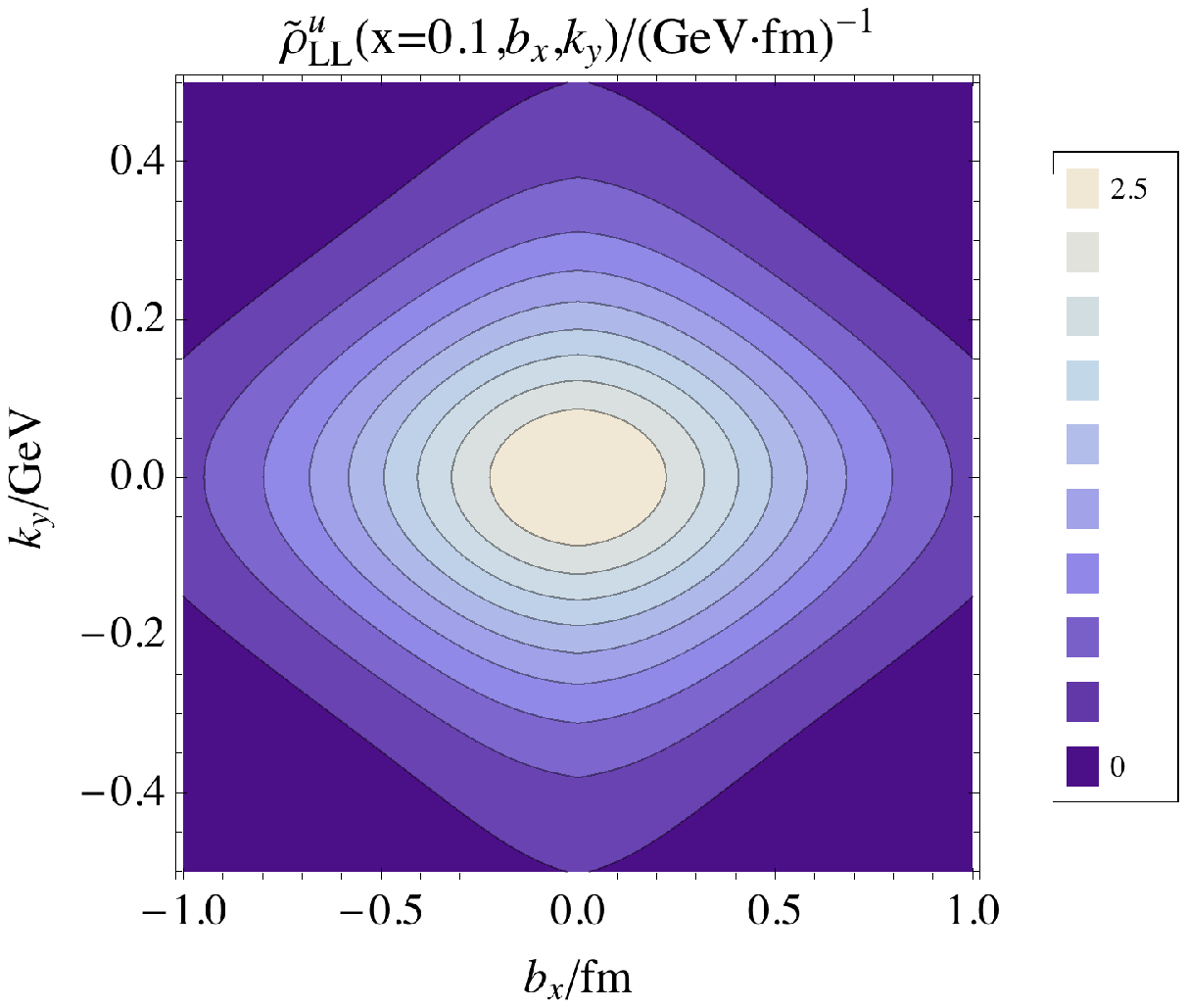}
\includegraphics[width=0.3\textwidth]{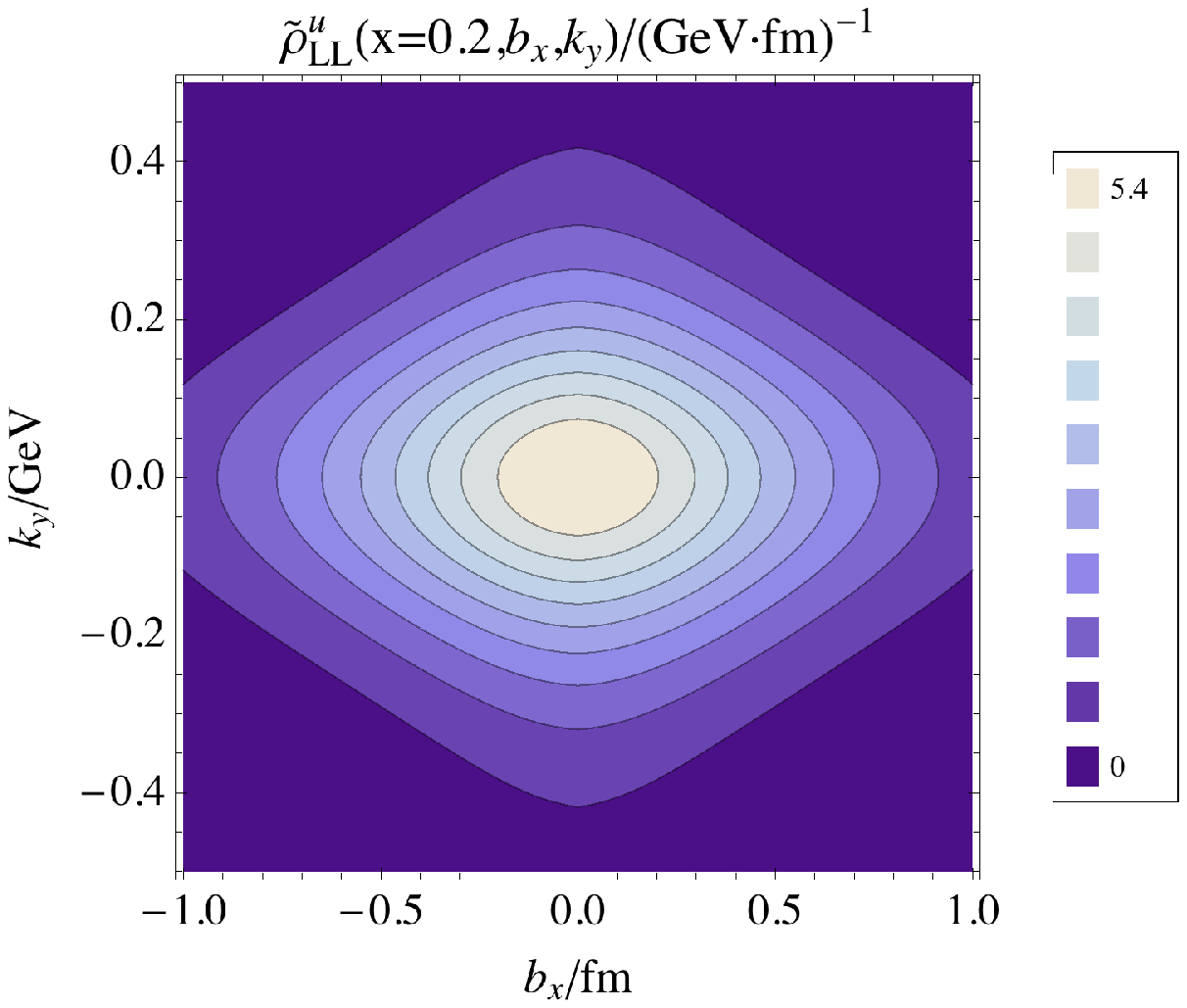}
\includegraphics[width=0.3\textwidth]{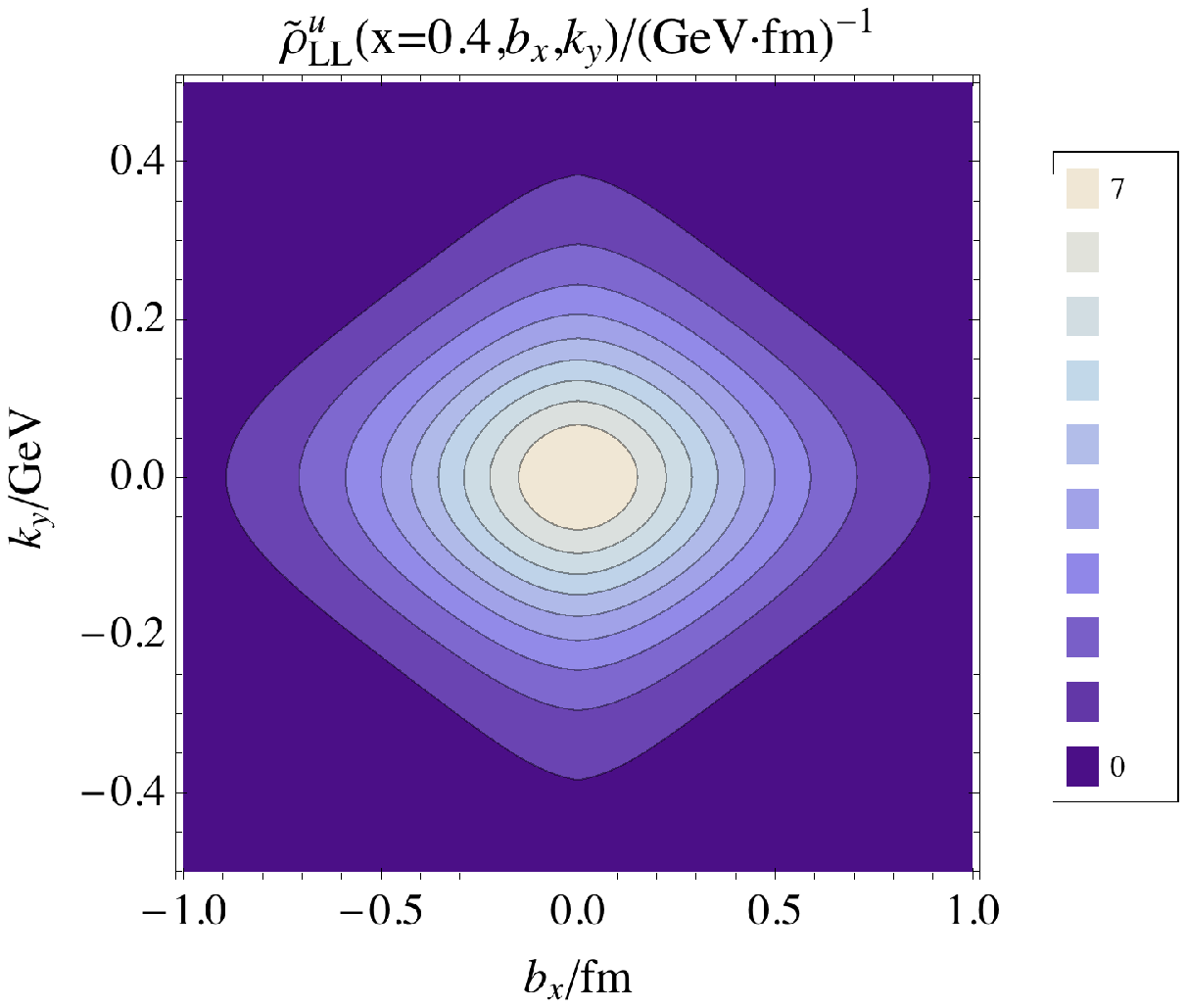}
\includegraphics[width=0.3\textwidth]{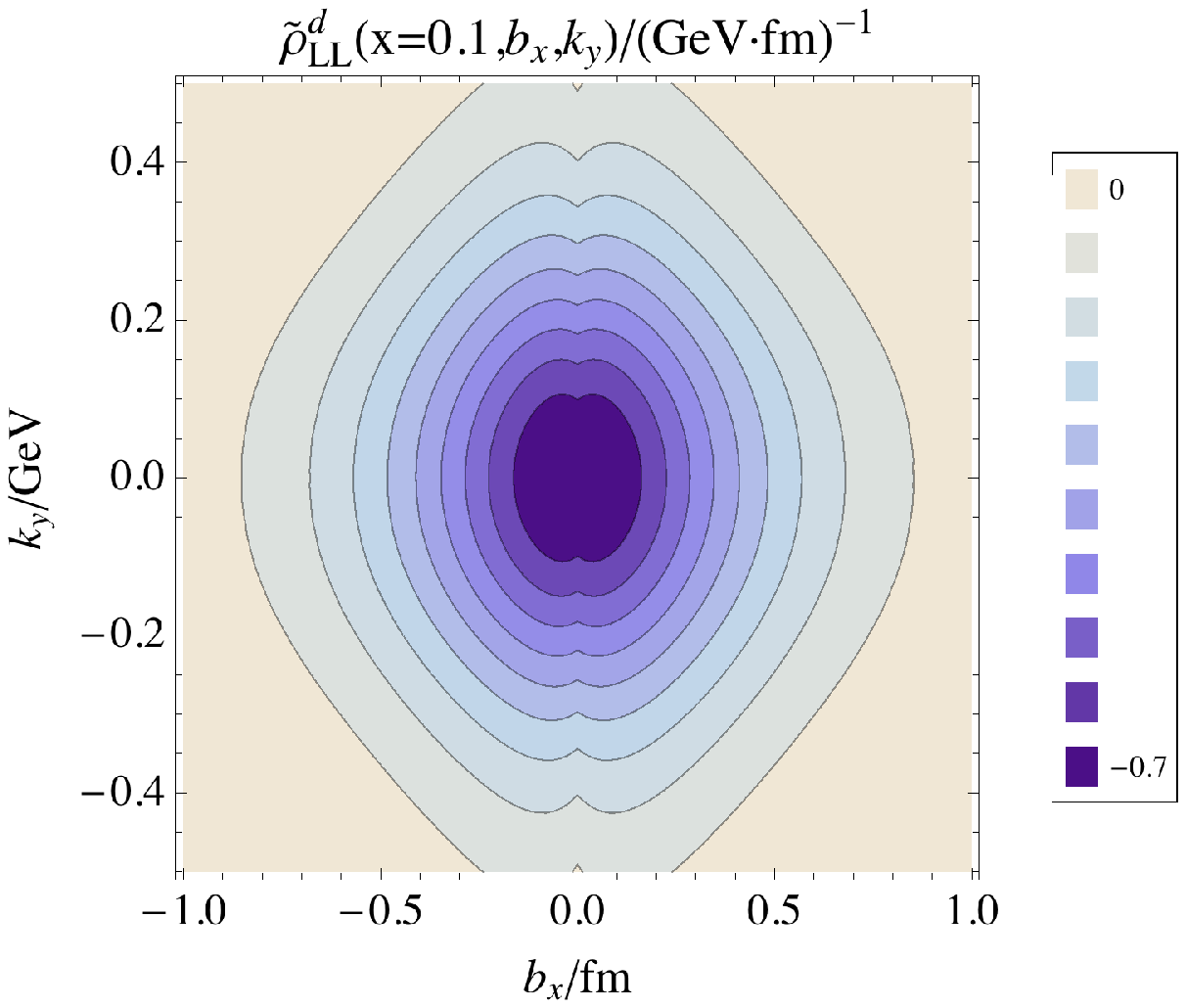}
\includegraphics[width=0.3\textwidth]{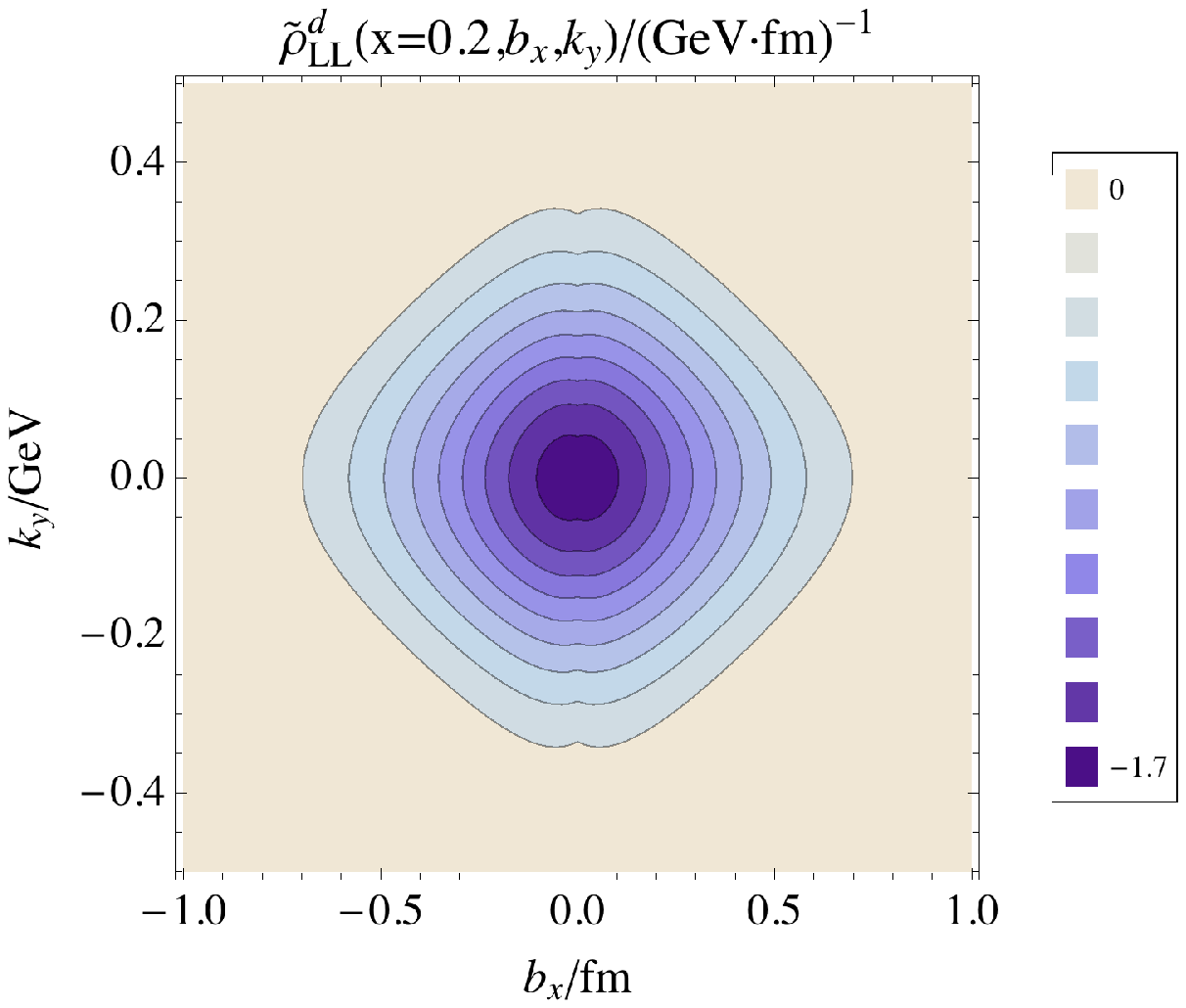}
\includegraphics[width=0.3\textwidth]{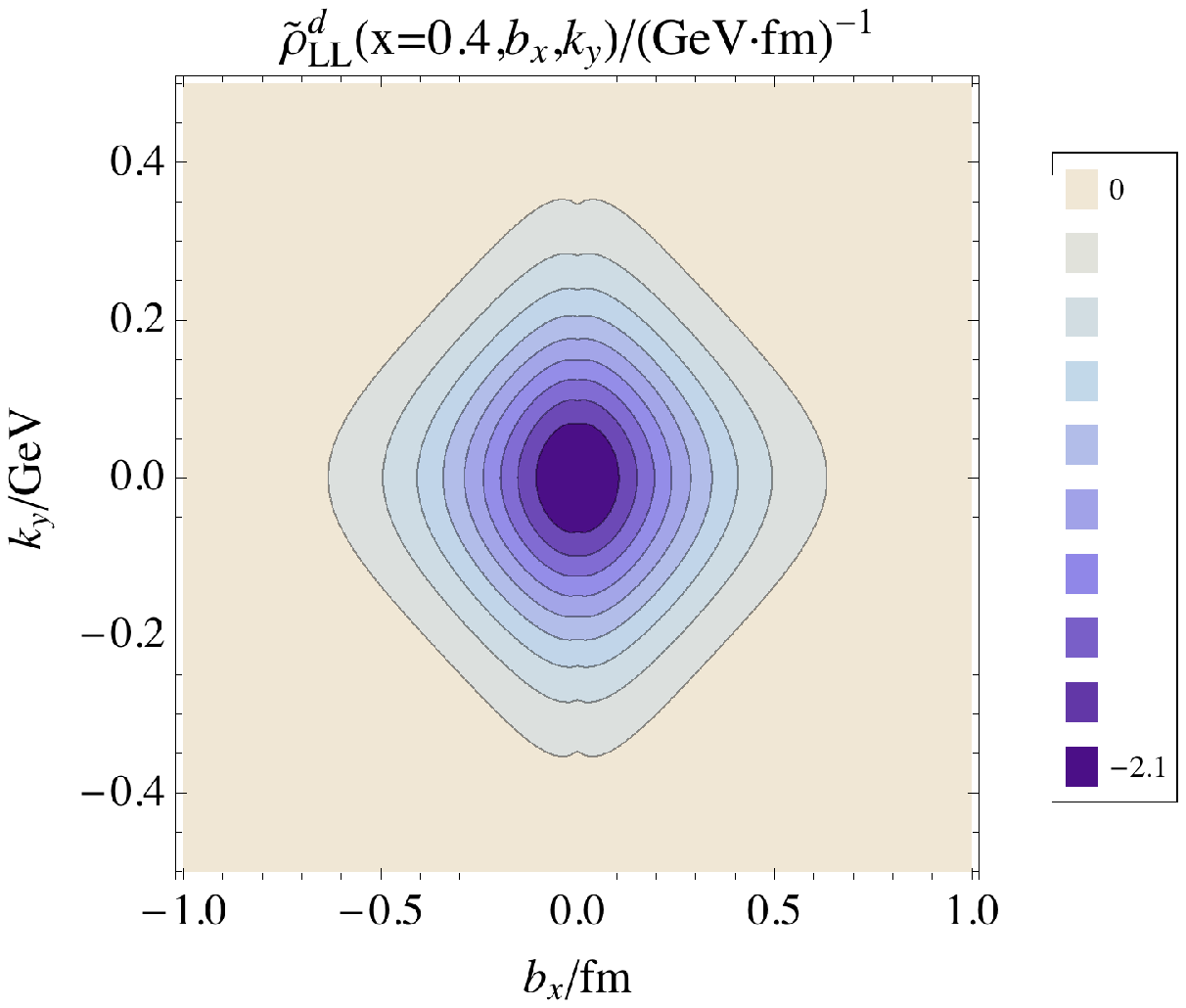}
\caption{(color online). Longitudinal mixing distributions $\tilde{\rho}_{_\textrm{LL}}(x,b_x,k_y)$ for $u$ quark (upper panels) and $d$ quark (lower panels) at $x=0.1$ (left column), $x=0.2$ (middle column) and $x=0.4$ (right column).\label{llmix}}
\end{figure}
\begin{figure}
\includegraphics[width=0.35\textwidth]{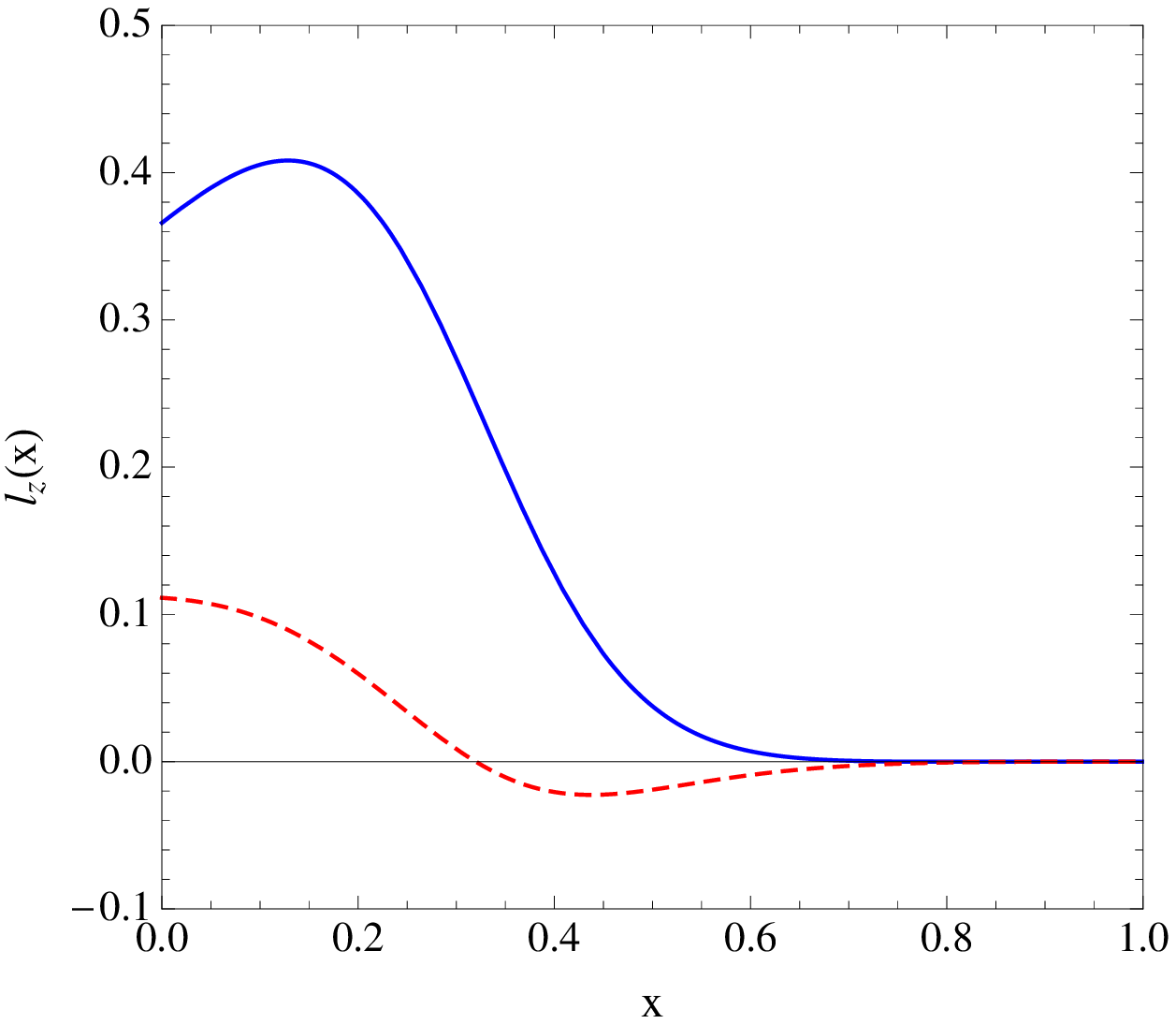}
\includegraphics[width=0.35\textwidth]{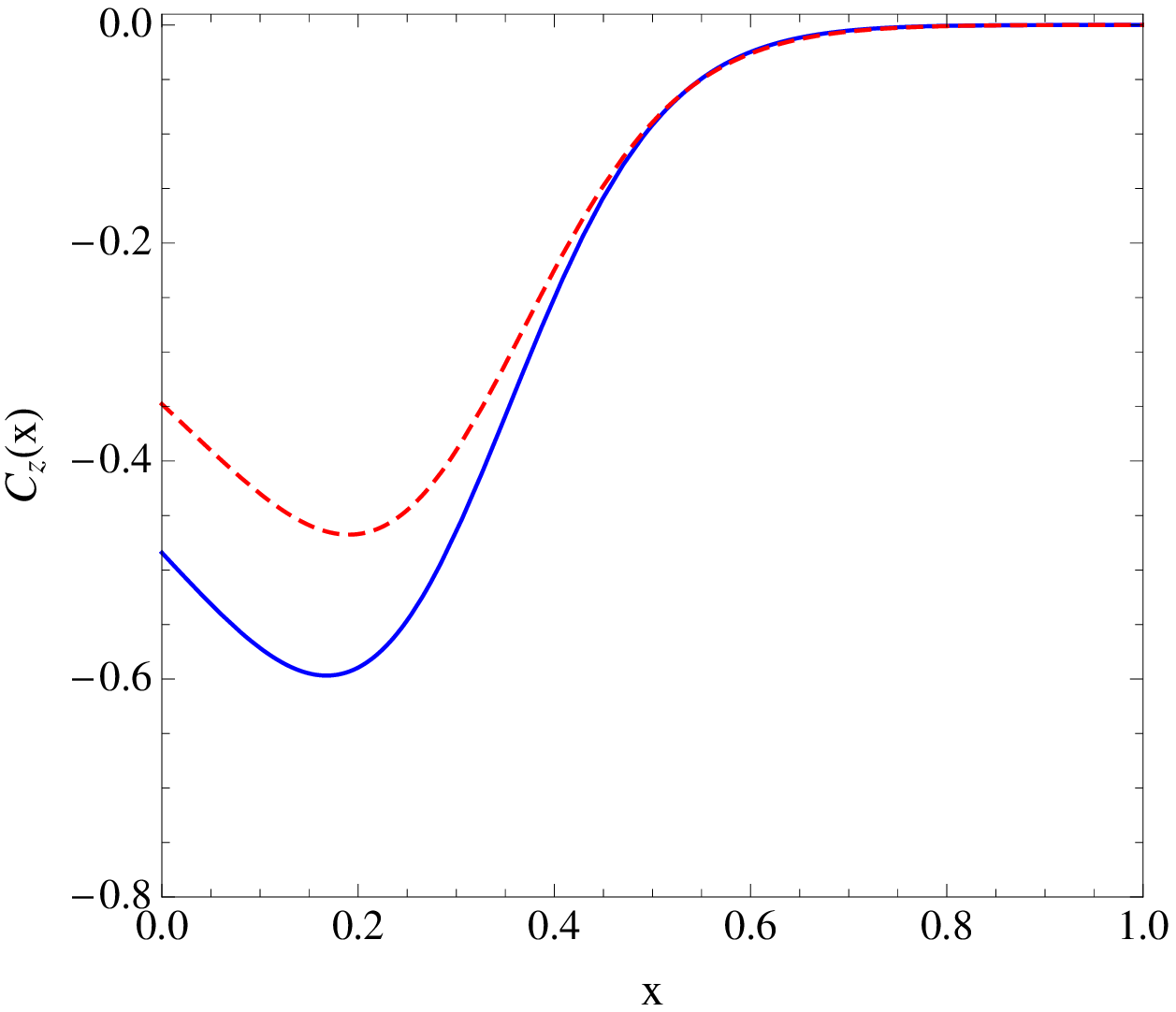}
\caption{(color online). Quark orbital angular momentum $\ell_z(x)$ (left) and spin-orbit correlator $\mathcal{C}_z(x)$ (right). The solid curve represents the $u$ quark and the dashed curve represents the $d$ quark.\label{lzcz}}
\end{figure}

In our model calculations, the masses $m$, $M_{s/v}$ and $M$ and the couplings $g_s$ and $g_v$ can be viewed as parameters to be fixed. One may also introduce some cutoff parameter $\Lambda$ to replace the quark mass $m$ in the denominators of light-cone wave functions. In this way, one will find the light-cone wave functions in~\cite{Jakob:1997wg} where the quark propagators are replaced by a dipole form factor with a cutoff mass $\Lambda$, and the corresponding expressions calculated in Sect. III are directly obtained by replacing the quark mass in the denominators in Eqs. (\ref{uu})-(\ref{pret}) and (\ref{uuv})-(\ref{pretv}) with the cutoff mass $\Lambda$. In this study, we will not do this replacement.

In principle, the spectator mass square $M_d^2$ has a spectrum. One may assume the shape of the spectrum or extract it from experimental data, but here we simply choose it as a parameter and fit it together with the quark mass parameter $m$ to the electomagnetic form factors. For the nucleon mass $M$, we adopt the value of proton mass in Ref.~\cite{Beringer:1900zz}. Then the coupling constants $g_s$ and $g_v$ are fixed by the quark number sum rules:
\begin{eqnarray}
\int dxd^2\bm{b}_\perp d^2\bm{k}_\perp\rho^u_{_\textrm{UU}}(x,\bm{b}_\perp,\bm{k}_\perp)&=&2,\\
\int dxd^2\bm{b}_\perp d^2\bm{k}_\perp\rho^d_{_\textrm{UU}}(x,\bm{b}_\perp,\bm{k}_\perp)&=&1.
\end{eqnarray}

With the light-cone wave functions, we can calculate Dirac and Pauli form factors from helicity-conserved and helicity-flip matrix elements of the plus component of the electromagnetic current operator~\cite{Brodsky:1980zm}:
\begin{eqnarray}
\left\langle P',\uparrow\left|\frac{J^+(0)}{2P^+}\right|P,\uparrow\right\rangle&=&F_1(Q^2),\\
\left\langle P',\uparrow\left|\frac{J^+(0)}{2P^+}\right|P,\downarrow\right\rangle&=&-\frac{\Delta^1-i\Delta^2}{2M}F_2(Q^2),
\end{eqnarray}
where
\begin{equation}
Q^2=-(P'-P)^2=\bm{\Delta}_\perp^2.
\end{equation}
The Sachs form factor are defined by the combinations of Dirac and Pauli form factors as~\cite{Sachs:1962zzc}
\begin{eqnarray}
G_E(Q^2)&=&F_1(Q^2)-\frac{Q^2}{4M^2}F_2(Q^2),\\
G_M(Q^2)&=&F_1(Q^2)+F_2(Q^2).
\end{eqnarray}

Then, the fitted values of the masses of the quark and the spectator are $m=451.4\,\textrm{MeV}$ and $M_d=705.3\,\textrm{MeV}$, and the results are plotted in Fig. \ref{ff}. The electromagnetic radius can be calculated from the derivatives of the form factors, and the values are $r_E^p=0.843\,\textrm{fm}$, $r_M^p=0.771\,\textrm{fm}$, $\langle r_E^{2}\rangle^n=-0.060\,\textrm{fm}^2$ and $r_M^n=0.703\,\textrm{fm}$.

The unpolarized, unpol-longitudinal, longi-unpolarized and longitudinal mixing distributions are shown in Figs. \ref{uumix}-\ref{llmix} respectively. They are plotted at three different $x$ values 0.1, 0.2 and 0.4 to show their $x$ dependence. The orbital angular momentum and spin-orbit correlator defined in (\ref{oam}) and (\ref{sl}) are shown in Fig. \ref{lzcz}. 

As observed from the unpolarized mixing distributions $\tilde{\rho}_{_\textrm{UU}}$ in Fig. \ref{uumix}, the left-right and top-bottom symmetries reflect that quarks have no preference to move either clockwise or anticlockwise at each point in coordinate and momentum spaces. This is consistent with the conclusion from the topology, because any preference will result in a privileged direction which breaks the isotropic property of the space.

As observed from the longi-unpolarized mixing distributions $\tilde{\rho}_{_\textrm{LU}}$ which represent the quark orbital motions in a longitudinal polarized proton in Fig. \ref{lumix}, we find that at small $x$ region both $u$ and $d$ quarks prefer moving anticlockwise and contribute positive orbital angular momentum, but at large $x$ region the $d$ quark distribution has a sign change and contribute negative orbital angular momentum. This is also observed in Fig. \ref{lzcz}.

The unpol-longitudinal mixing distributions $\tilde{\rho}_{_\textrm{UL}}$ represent the correlation between the quark spins and orbital motions. As observed in Fig. \ref{ulmix}, both $u$ and $d$ quarks prefer moving clockwise which means negative correlations between quark spins and orbital motions. However, this does not necessarily result in opposite signs of the quark spin and the orbital angular momentum. We take the scalar spectator case as an example. In a positive polarized proton, we have two quark-scalar-spectator spin states. One has a positive polarized quark with zero orbital angular momentum and has no contributions to the spin-orbit correlator. The other one has a negative polarized quark with positive orbital angular momentum and has negative contributions to the spin-orbit correlator. Thus, if the first one plays a dominant role, we will find both positive quark spin and positive orbital angular momentum but negative spin-orbit correlator in average.

\section{Conclusions}

We investigated quark Wigner distributions in a spectator model. With the combinations of the polarization configurations of the quark and the nucleon, ten independent Wigner distributions are defined. We calculated all these distributions in the model with both the scalar and the axial-vector spectators. In our calculations, we derived the light-cone wave functions with effective quark-spectator-nucleon interactions, and then generalized the perturbative amplitudes by adjusting the power of the energy denominators. In this procedure, we respected the Lorentz invariance, and as seen from the explicit expressions, they are frame independent.

In order to include the effects of the gauge link in Wigner operator, we introduced a phase to each light-cone amplitude. We estimated the phases from the one gluon exchange interactions. Though each of them is infrared divergent, the differences between them are infrared finite, and so we only keep the relative phases. Then, we calculated all the Wigner distributions with the light-cone wave functions and the relative phases from the gauge link. Some relations are found from the expressions of the distributions. By integrating over transverse coordinates or transverse momenta, they will reduce to the relations between TMDs or IPDs. We also calculated the orbital angular momentum $\ell_z$ and $\mathcal{C}_z$ defined in the literature~\cite{Lorce:2011kd}, and kept the longitudinal momentum fraction unintegrated to represent their $x$ dependence.

Apart from TMDs and IPDs, one may also define the mixing distributions by integrations over one transverse coordinate and one transverse momentum along two orthogonal directions. Unlike the Wigner distributions, the mixing distributions have the probability interpretations and describe the correlation between quark transverse coordinate and transverse momentum. Therefore, quark orbital motions are clearly seen from mixing distributions. The parameters were fitted from the electromagnetic form factors. We only provided the numerical results of the mixing distributions for unpolarized and longitudinal polarized quark or proton, but with the values of parameters and expressions of Wigner distributions one may easily get the results for transverse polarized quark or proton. In a polarized proton, both $u$ quark and $d$ quark prefer moving anticlockwise and contribute positive orbital angular momentum, but at large $x$ region a sign change was observed for the $d$ quark orbital motions.

\acknowledgements{The author thanks Barbara Pasquini, Alessandro Bacchetta and Marco Radici for helpful discussions. The author also acknowledges the host by INFN, Pavia, where this work was finished.}


%

\end{document}